\documentclass{ws-ijmpa}
\usepackage[super,compress]{cite}
\usepackage{graphicx}
\begin{document}
\markboth{Wilk and W\l odarczyk}{Counting distributions from the perspective of combinants}

\catchline{}{}{}{}{}
%

\title{Counting distributions from the perspective of combinants}

\author{Grzegorz Wilk}

\address{National Centre for Nuclear Research, Department of Fundamental Research,\\ Pasteura 7, St. 02-093 Warsaw, Poland\\
grzegorz.wilk@ncbj.gov.pl}

\author{Zbigniew W\l odarczyk}
\address{Institute of Physics, Jan Kochanowski University,\\ Uniwersytecka 7 St., 25-406 Kielce, Poland\\
zbigniew.wlodarczyk@ujk.kielce.pl}

\maketitle

\begin{history}
\received{Day Month Year}
\revised{Day Month Year}
\end{history}

\begin{abstract}

We present a comprehensive insight into counting distributions from the perspective of the combinants extracted from them. In particular, we focus on cases where these combinants exhibit oscillatory behavior that can provide an invaluable new source of information about the dynamics of the process under study. We show that such behavior can be described only by specific combinations of compound distributions based on the Binomial Distribution and provide their analytical forms which can be used in further investigations and which can be helpful in the analysis of all other types of counting distributions.

\keywords{multiparticle production; statistical models; counting distributions, combinants}
\end{abstract}

\ccode{PACS numbers: 13.85.Hd, 25.75.Gz, 02.50.Ey}


\section{Introduction}
\label{sec:Introduction}

A counting distribution is a discrete distribution with only non-negative integers in its domain. We have to deal with them wherever we encounter sequences of identical events, write down the distributions of these cases in the form of certain histograms, and finally, analyze them and try to determine what known multiplicity distributions they correspond to, i.e. to find the mechanism of their formation.

In this work we will present a comprehensive insight into these distributions using the example of multiparticle production processes. In this case the experimentally measured multiplicity distributions $P(N)$ of the produced particles (our measured counting distributions) are regarded as one of the first important sources of information about the dynamics of the multiparticle production processes. The details of their shapes contain invaluable information about this dynamics, obtained by examining the different characteristics of these distributions, for example factorial moments and cumulant factorial moments (or their ratios) \cite{Kittel,Book-BP}. They are very sensitive to the details of the multiplicity distribution because of their oscillatory behavior as a function of the rank $q$ which, in turn, seems to follow theoretical predictions. However, like all moments, they require knowledge of all multiplicities $P(N)$ and are therefore very sensitive to any constraints on the available phase space (which themselves can be the source of the observed oscillations). Such a limitation is not found in other characteristics of $P(N)$ distributions, such as combinants, which depend only on multiplicities smaller than their rank \cite{ST,Combinants-1,Combinants-2}. They are the topic of our work which continues our previous investigations presented in \cite{Ours1,WW-S,RWW-S,Ours2,Ours3,Ours-Odessa,Ours-EPJA}.

In the literature, combinants appear in two different (although equivalent) ways. They can appear as coefficients in the recurrence relation defining $P(N)$ which is, for example, widely explored in counting statistics when dealing with multiplication effects in point processes \cite{ST},
\begin{equation}
(N + 1)P(N + 1) = \langle N\rangle \sum^{N}_{j=0} C_j P(N - j). \label{rr}
\end{equation}
Here the coefficients $C_j$ contain the memory of particle $N+1$ about all the $N-j$ previously produced particles. By inverting the recurrence relation (\ref{rr}) one obtains a recurrence relation for $C_j$,
\begin{equation}
\langle N\rangle C_j = (j+1)\left[ \frac{P(j+1)}{P(0)} \right] - \langle N\rangle \sum^{j-1}_{i=0}C_i \left[ \frac{P(j-i)}{P(0)} \right], \label{rCj}
\end{equation}
which allows direct determination of $C_j$ from the experimentally measured multiplicity distribution $P(N)$.
We first used this approach in \cite{Ours1,WW-S} where we proposed to call the coefficients $C_j$ {\it modified combinants} (to distinguish them from the previously introduced, in a different way described below, combinants $C^{\star}_j$; both coefficients are closely related to each other). It turns out that  the $C_j$ obtained from high energy LHC experiments (provided they have sufficiently large statistics) show distinct oscillatory behavior with rank $j$ \cite{Ours1} which cannot be reproduced by a single NBD commonly used to fit data (as shown later in \cite{Zborovsky}, for this you need some suitably weighted sum of NBDs). This behavior of the coefficients $C_j$ deduced from the measured distributions $P(N)$ will now be the subject of further research.

As already mentioned, the modified combinants $C_j$ are closely related to the {\it combinants} $C^{\star}_j$ introduced earlier in \cite{Combinants-1,Combinants-2} by means of the generating functions $G(z)$ of the multiplicity distributions $P(N)$,
\begin{equation}
G(z) = \sum_{N=0}^{\infty} P(N) z^N \label{GF}
\end{equation}
such that
\begin{equation}
P(N) = \frac{1}{N!}\frac{d^N G(z)}{d z^N}\bigg|_{z=0} \label{PN}
\end{equation}
and
\begin{equation}
C^{\star}_j = \frac{1}{j!}\frac{d^j \ln[G(z)]}{d z^j}\bigg|_{z=0}. \label{lnGF}
\end{equation}
As shown in \cite{Combinants-2,VVP} the $C^{\star}_j$ completely characterise the probabilities $P(N)$ because each $C^{\star}_j$ can be expressed as a finite combination of the first $j$ probability ratios $\left[ P(j)/P(0)\right]$ (this is why they were called combinants). This property distinguishes combinants from factorial moments $F_q$ and cumulant factorial moments $K_q$, defined as
\begin{equation}
F_q = \frac{d^q G(z)}{d z^q}\bigg|_{z=1} = \sum_{N=q}^{\infty}\frac{N!}{(N-q)!}P(N)\label{FqfromG}
\end{equation}
and
\begin{equation}
K_q = \frac{d^q \ln G(z)}{d z^q}\bigg|_{z=1} = \sum_{j=q}^{\infty}\frac{j!}{(j-q)!}C^*_j, \label{KqfromG}
\end{equation}
which also characterize $P(N)$, but which require knowledge of all terms of the distribution $P(N)$ (note that whereas $P(N)$ defines $F_q$ the $C^*_j$ define $K_q$). Consequently, while the moments $F_q$ and $K_q$ are suited to study the densely populated region of phase space, combinants $C^{\star}_j$ are better suited for the sparsely populated regions \cite{Kittel,Book-BP}. On the other hand, combinants share with them their additivity property. Namely, for a random variable composed of independent random variables, with its generating function given by the product of their generating functions, $G(x)=\prod_jG_j(x)$, the corresponding modified combinants are given by the sum of the independent components. All of this remains true for modified combinants $C_j$ because \cite{Ours1}
\begin{equation}
C_j = \frac{j+1}{\langle N\rangle} C^{\star}_{j+1},\qquad {\rm where}\qquad \langle N\rangle = \sum_{N=0}^{\infty} N P(N). \label{CstarC}
\end{equation}

In conclusion, this approach seems to be more useful if we want to determine what form of phenomenological distribution of multiplicities most faithfully reproduces both the measured $P(N)$ and the $C_j$ (or $C^{\star}_j$) obtained from it. Apart from \cite{Combinants-1,Combinants-2,VVP} and our works \cite{Ours1,WW-S,RWW-S,Ours2,Ours3,Ours-Odessa,Ours-EPJA}, it has been used in a number of other publications \cite{Kittel,PCarruthers,Book-BP,BS,B-AIP,SzSz,LeeM,MCH,Hegyi-1,Hegyi-2,Hegyi-3}. However, only in \cite{Ours1,WW-S,RWW-S,Ours2,Ours3,Ours-Odessa,Ours-EPJA,BS,B-AIP,Hegyi-1,Hegyi-2,Hegyi-3} was the possibility of their oscillatory behavior for $P(N)$ measured in high energy collisions noticed and investigated.

Whereas in the works \cite{Hegyi-1,Hegyi-2,Hegyi-3} oscillations were taken as evidence that the investigated multiplicity distributions  $P(N)$ are not infinite divisible \cite{Book-BP}, here, based on our results obtained so far \cite{Ours1,WW-S,RWW-S,Ours2,Ours3,Ours-Odessa,Ours-EPJA}, we want to argue that they signal that the dynamics of the particle production process is of such a type that it should be described only by some compound distribution (CD) based on the Binomial Distribution (BD). This fact has to be contrasted with the completely different behavior of combinants obtained from PD, which are always non-oscillatory (the most popular example is the very popular Negative Binomial Distribution (NBD) which is a CD of a PD and a logarithmic distribution \cite{Kittel,Book-BP}).

In the next Section we will recall the most important characteristics of the BD, PD and NBD, the three basic distributions used in various combinations in the descriptions of the measured $P(N)$. In particular, we will describe the already mentioned specific role of the BD in these descriptions. In Section \ref{CDinBP} we will show in detail how, using Bell polynomials, one can describe the compound distributions $P_{CD} = P_{BD}\circ P_{PD}$ and $P_{CD} = P_{BD}\circ P_{NBD}$ in a compact analytical way and we will estimate the period of the oscillations of $C^{\star}_N$. This problem is further illustrated in Section \ref{Toy-model} by using a simple toy model to obtain analytical results. The analytical calculation method used leads to clear formulas for the number distributions and the corresponding combinants, which depend only on some well-defined combinations of parameters from the distributions contained in the CDs under consideration and on well-defined, though quite complex, combinatorial factors (described by Bell polynomials). These results can explain how the numerical algorithms used to fit the experimental data work and, in particular, should help find the best combinations of the parameters used. Section \ref{OtherCMD} shows two examples of other possible types of combined multiplicity distributions, the weighted sum of multiplicity distributions and  the sum of the multiplicity distributions of different types, and their relation with results presented in  Section \ref{CDinBP}. In Section \ref{Inter} for completeness of our presentation we present, using Bell polynomials, the interconnections of multiplicity distributions and combinants. Section \ref{SumDisc} summarizes our work and includes the conclusions drawn from it.
 ll the necessary formulas used in the work can be found in \ref{AddInf} whereas derivations of results presented in Sections \ref{CD=BDxPD} and \ref{CD=BDxNBD} are presented, respectively, in \ref{Derivation-1} and \ref{Derivation-2}.

\section{Setting the stage}
\label{Survey}

To set the stage we present the most important characteristics of the multiplicity distributions considered: the Binomial (BD), Poisson (PD) and Negative Binomial (NBD).

The Binomial Distribution appears to be a type of fundamental distribution in the sense that it is the only one possible from the point of view of conservation rules\footnote{For example, only for this distribution can one have energy conservation for the sum of the energies of the particles in some energy distribution for a given number of particles, $N$. For $N=const$ we have that $P(E) \sim (1 - E/U)^N$. However, if the particles are taken from some distribution $P(N)$ then only for a BD we will have particles whose energy will not exceed $U$. For comparison, for $P(N)$ in the form of a Poisson Distribution (PD) we have an exponential energy distribution in which all energies are possible.}. It is defined by the generating function:
\begin{equation}
G(z) = [ 1 - p(1 - z)]^K = \left[ 1 - \frac{\langle N\rangle}{K}(1-z)\right]^K,\qquad p = \frac{\langle N\rangle}{K}. \label{G-BD}
\end{equation}
Changing $K$ to $-k$ in the BD, we get the NBD with the generating function\footnote{There are many scenarios leading to the NBD, the most important from our point of view are: fluctuations in the $\lambda$ parameter in the PD given by the gamma distribution, the composition of $k$ geometric distributions, and, finally, the composition of a PD with a logarithmic distribution.}
\begin{eqnarray}
G(z) = \left( \frac{1 - p}{1 - pz}\right)^k &=& \left[ 1 - \frac{p}{1-p}(z-1)\right]^{-k} =\nonumber\\
&=& \left[ 1 - \frac{\langle N\rangle}{k}(z-1)\right]^{-k},\quad p = \frac{\langle N\rangle}{\langle N\rangle +k}.\label{G-NBD}
\end{eqnarray}
In the limit $K \to \infty$ with $\langle N\rangle = const$ the BD becomes the PD with generating function
\begin{equation}
G(z) = e^{\lambda (z-1)},\qquad \lambda = \langle N\rangle, \label{G-PD}
\end{equation}
in the case of the NBD this happens in the limit $k \to \infty$ and $\langle N\rangle = const$,

In Table \ref{Table_1} we present a comparison of these distributions. Their common feature is that they are described by at most two parameters that can be expressed through the first two moments of these distributions. Therefore, the corresponding combinants do not contain more information than the first two moments of the probability distribution,
\begin{equation}
C_j = \frac{\langle N\rangle}{Var(N)}\left( 1 - \frac{\langle N\rangle}{Var(N)}\right)^j. \label{TwoMoments}
\end{equation}
Let us note that the possibility of the occurrence of an oscillation is inseparable from the fact that in Eq. (\ref{TwoMoments}) we must have $Var(N) < \langle N\rangle$, which in this case is a characteristic distinguishing feature of the BD.
For these distributions $P(N)$, the recursive formula (\ref{rr}) reduces to a simple form
\begin{equation}
(N + 1)P(N+1) = g(N) P(N)~{\rm where}~ g(N) = \frac{\langle N\rangle^2}{Var(N)} + \left[ 1 - \frac{\langle N\rangle}{Var(N)} \right] N. \label{Simplerr}
\end{equation}

\begin{table}[h]
\tbl{Distributions $P(N)$ used in this work: Binomial (BD), Poisson (PD) and Negative Binomial (NBD),  their generating functions $G(z)$ and the modified combinants $C_j$ emerging from them, mean multiplicities $\langle N\rangle$ and their variances $Var(N)$. }
{\begin{tabular}{@{}llcccc@{}} \toprule
                            & $P(N)$                                     & $G(z)$   & $C_j$   & $\langle N\rangle$  & $Var(N)$     \\ \colrule
BD                          & ${ K\choose N} p^N (1 - p)^{K-N}$          & $(pz + 1 - p)^K$   & $(-1)^j\frac{K}{\langle N\rangle} \left( \frac{p}{1-p}\right)^{j+1}$   &  $Kp$                & $\langle N\rangle - \frac{\langle N\rangle^2}{K}$        \\
PD                          & $\frac{\lambda^N}{N!} \exp( - \lambda)$    & $\exp[\lambda (z - 1)]$      & $\delta_{j,0}$    & $\lambda$                & $\langle N\rangle$         \\
NBD                         & $\binom{N+k-1}{N} p^N (1 - p)^k$           & $\left( \frac{1 - p}{1 - pz}\right)^k$   & $\frac{k}{\langle N\rangle} p^{j+1}$    & $\frac{kp}{1-p}$                & $ \langle N\rangle + \frac{\langle N\rangle^2}{k}$        \\ \botrule
\end{tabular} \label{Table_1}}
\end{table}

The experimentally observed distributions $P(N)$ lead to combinants that do not satisfy (\ref{TwoMoments}), namely, we either observe oscillations with period $2$, i.e. the same as for the BD, but $Var(N) > \langle N\rangle$, or we have oscillations with a period greater than $2$. In such cases, we either refer to single distributions that are certain generalized forms of distributions from Table \ref{Table_1} (described by more than $2$ parameters) \cite{Ours1,Hegyi-2} or to a different type of distribution composed of those presented in Table \ref{Table_1} (which consequently also leads to a greater number of parameters introduced into the description). In the next three Sections we will consecutively consider examples of this type of distribution.

\section{Compound Distributions}
\label{CDinBP}

\subsection{Introduction and notation}
\label{DerCompDist}

The need for compound  distributions (CD) always arises whenever individual distributions are unable to describe the measured multiplicity distributions. In such cases one usually assumes that the production process is two-stage, with a number $M$ of some objects (clusters/fireballs/etc.) produced according to some distribution $f(M)$ (defined by a generating function $F(z)$), which subsequently decay independently into a number of secondaries, $n_{i = 1,\dots, M}$, following some other (always the same for all $M$) distribution, $g(n)$ (defined by a generating function $G(z)$). The resultant multiplicity distribution $P(N)$ is a compound distribution of $f(M)$ and $g(n)$, $P(N) = f(M)\circ g(n)$, with generating function $H(z)$, where
\begin{equation}
N = \sum_{i=1}^M n_i,\qquad P(N) = \sum_M g(n|M)f(M)\qquad {\rm and}\qquad H(z) = F[G(z)]. \label{CD-all}
\end{equation}
For such distributions
\begin{equation}
\langle N\rangle = \langle M\rangle \langle n\rangle\qquad {\rm and}\qquad Var(N) = \langle M\rangle Var(n) + \langle n\rangle^2 Var(M).\label{CD-mean-Var}
\end{equation}

We shall present detailed analytical derivations of the compound distributions $P(N)$ and their combinants $C^{\star}_N$ used by us, i.e. the compound distributions of PD and NBD based on the BD: $P_{CD} = P_{BD}\circ P_{PD}$ and $P_{CD} = P_{BD}\circ P_{NPD}$. The basic steps are, respectively, the corresponding derivatives of $H(z) = F[G(z)]$ and $L(z) = \ln H(z)$. They can be calculated using one of the following two forms; to calculate $P(N)$ we need:
\begin{eqnarray}
\frac{d^N H(z)}{d z^N} &=& \sum_{\left\{r_1,r_2,\dots,r_N \right\}} F^{(r)}[G(z)] \left[ \frac{N!}{r_1! r_2 ! \cdots r_n!} \right] \prod_{m=1}^N \left[ \frac{G^{(m)}(z)}{m!}\right]^{r_m} =\label{FdiB}\\
&=& \sum_{r=1}^N F^{(r)}[G(z)]\cdot B_{N,r}\left[ G^{(1)}(z),G^{(2)}(z),\dots,G^{(N-r+1)}(z)\right], \label{BellP}
\end{eqnarray}
\begin{eqnarray}
B_{N,r}\left[ G^{(1)},\dots,G^{(N-r+1)}\right] &=& \sum_{\left\{r_1,r_2,\dots,r_N \right\}} \left[ \frac{N!}{r_1! r_2 ! \cdots r_N!} \right] \prod_{m=1}^{N-k+1} \left[ \frac{G^{(m)}(z)}{m!}\right]^{r_m}, \label{FDB}\\
F^{(r)}(G) &=& \frac{d^r F(G)}{d G^r}\qquad {\rm and}\qquad G^{(m)}(z) = \frac{d^m G(z)}{d z^m}. \label{defHG}
\end{eqnarray}
Eq. (\ref{FdiB}) is the Fa\`a di Bruno formula \cite{FdB,Johnson,AroundFdB}), the sum is over all $\left( r_1,r_2,\dots,r_N\right) = \left\{ r_i\right\}$ such that
\begin{equation}
r_1 + 2r_2 + \cdots + N r_N = N\qquad{\rm and}\qquad r_1 + r_2 + \cdots + r_N = r. \label{Sumrs}
\end{equation}
After combining terms with the same value of $r_1 + r_2 + \cdots + r_n = r$ and noting that $r_j = 0$ for $j > N-r+1$, Eq. (\ref{FdiB}) can be rewritten in terms of partial Bell polynomials, Eq. (\ref{BellP}) (see Section \ref{BP}). The same procedure applies to $L(z)$ in the case of combinants.

\subsection{Compound PD}
\label{NBDasCD}

Before proceeding to the actual calculations, let us come back for a moment to the problem of the source of the combinants' oscillations. To this end, we will briefly look at two simple examples of CD: the Poisson compound distribution (PCD) and the NBD as a combined distribution of a PD and a logarithmic distribution.

Note that the combinants $C^{\star}_j$ allow any $P(N)$ distribution to be represented as a departure from the Poisson distribution \cite{Combinants-1,Combinants-2,VVP}, one only needs to replace the first degree polynomial in the logarithm of the PD generating function with a power series, the coefficient of which are the combinants $C^{\star}_j$:
\begin{equation}
G(z)= \sum_{N=0}^{\infty}z^N P(N) = \exp[\langle N\rangle  (z-1)]\quad \Rightarrow\quad \ln[G(z)] = \ln[P(0)] + \sum_{j=1}^{\infty} C_j^{\star} z^j \label{CstarJN}
\end{equation}
under the condition  that $P(0)>0$. Normalization $\sum_{N=0}^{\infty} P(N)=1$ implies that $G(1)=1$, therefore
\begin{equation}
\ln[P(0)] = - \sum_{j=1}^{\infty} C_j^{\star} \quad \Longrightarrow\quad G(z) = \exp\left[ \sum_{j=1}^{\infty} C_j^{\star}\left(z^j - 1\right)\right]. \label{BF-1}
\end{equation}
 This means that the combinants $C_j^{\star}$ fully characterize the multiplicity distribution $P(N)$. The simplest case is the PCD with generating function
\begin{equation}
H(z) = \exp\left\{ \lambda [ G(z) - 1]\right\}, \label{infdiv}
\end{equation}
where $\lambda > 0$ and $G(z)$ is another probability generating function. Note that the combinants derived from any PCD show no oscillations because in this case
\begin{equation}
C^{\star}_j = \frac{\lambda}{j!} \frac{d^j G(z)}{d z^j}\bigg|_{z=0} = \lambda g(j), \label{Cjcalculated}
\end{equation}
where $g(j)$ is the multiplicity distribution obtained from the generating function $G(z)$. Since it is strictly positive, the resulting combinants are also strictly positive. On the other hand, a generating function of the form of Eq. (\ref{infdiv}) guarantees that the corresponding distribution is infinitely divisible \cite{InfDiv,InfDiv-1}. Therefore, the common criterion for the absence or appearance of oscillations is whether or not a given distribution is infinitely divisible \cite{Book-BP,Hegyi-1,Hegyi-2,Hegyi-3}. Note that, for example, PCD can be regarded as convolutions of Poisson singlet, doublet, triplet, etc. distributions because Eq. (\ref{infdiv}) can also be written as the product of the generating functions of the Poisson components having mean values $C^{\star}_q$,
\begin{equation}
G(z) = \prod_{q=1}^{\infty}\exp\left[ C^{\star}_q \left( z^q - 1\right)\right]. \label{infdiv-1}
\end{equation}
Because the $C^{\star}_q$ are proportional to the probabilities given by $g(z)$ with a constant of proportionality $\lambda$, hence for infinitely divisible distributions they cannot be negative \cite{InfDiv}.

This property of PCD explains why the NBD combinants do not oscillate. It is because it itself is a compound distribution of a PD (with generating function (\ref{G-PD})) and a logarithmic distribution (LD), $P_{NBD} = P_{PD}\circ P_{LD}$, where
\begin{equation}
P_{LD} = -\frac{p^N}{N\ln(1-p)}\qquad {\rm and~its~generating~function}\qquad G(z) = \frac{\ln(1-pz)}{\ln(1-p)}. \label{Gfs-LD}
\end{equation}
The generating function of such a distribution is therefore
\begin{equation}
H(z) = \exp\left\{ \lambda[G(z)-1]\right\} = \frac{e^{-\lambda}}{(1 - pz)^L}\quad{\rm where}\quad L = \frac{-\lambda}{\ln(1-p)}, \label{CPD=PD&LD}
\end{equation}
which for $L =k$ results in the generating function of the NBD given in Eq. ({\ref{G-NBD}) \footnote{Note that while it is possible to obtain from a PD a wider NBD distribution, it proves impossible to obtain from a PD a narrower distribution, such as a BD. Such a procedure would require that $G_{\tiny{BD}}(z) = e^{\lambda \left[F_X(z)-1\right]} \Longrightarrow F_X(z) =\lambda + \frac{1}{\lambda}\ln G_{\tiny{BD}}(z)$ and that $P_X(N) = \frac{1}{N!}\frac{d^N G_{\tiny BD}(z)}{d z^N}\bigg|_{z=0} = \frac{1}{\lambda} C_N^{\star}(BD)$, i.e., the respective multiplicity distribution needed for this purpose would have to be strongly oscillatory functions of $N$, following the oscillatory behavior of the BD combinants.}.

We would now like to note that such an apparently firm association of an infinite divisibility of $P(N)$ with oscillations  of the combinants is too strong because it is possible to give a counterexample of a distribution which is not infinitely divisible but whose $C_j^{\star}$ s are strictly positive. It is a uniform distribution,
\begin{equation}
P(N)=\frac{1}{K+1},~ N\in [0,K],\qquad {\rm with}\qquad G(z)= \frac{1 - z^{K+1}}{(K+1)(1-z)},\label{UD}
\end{equation}
which is not infinitely divisible and for which the resulting $C_j$ are strictly positive, in fact from Eq. (\ref{rCj}) with $P(j)/P(0) = 1$ one immediately gets that
\begin{equation}
C_j = \frac{1}{\langle N\rangle} = \frac{2}{K}\qquad {\rm and,~respectively,}\qquad C^{\star}_{j+1} = \frac{1}{j+1}. \label{UD-C}
\end{equation}
This means that a lack of oscillations does not necessarily mean that the given distribution is infinitely divisible.

Therefore, in what follows we shall focus on the observation that oscillations of combinants always appear where the $P(N)$ is either of the pure BD type, or a compound distribution of the BD with some other distribution (which can itself be again any other compound distribution) \cite{Ours2,Ours3,Ours-Odessa,Ours-EPJA}.

\subsection{Compound distribution $P_{CD} = P_{BD}\circ P_{PD}$}
\label{CD=BDxPD}

Using $F(z)$ and $G(z)$ for, respectively, BD and PD as given in Table \ref{Table_1} we have that the resulting multiplicity distribution $P(N)$
can be presented in one of the following forms (details of their derivation are presented in \ref{Derivation-1}):
\begin{eqnarray}
P(N) &=& \frac{\lambda^N}{N!} X^K \sum_{r=0}^{min(N,K)} r! \binom{K}{r} S(N,r) Y^r =\label{PPD-restr-1}\\
&=& \frac{\lambda^N}{N!}\left( 1 + p'e^{-\lambda}\right)^K\sum_{r=0}^{min(N,K)} P_{BD}(r)\cdot \left[ r! e^{-r\lambda} S(N,r)\right]. \label{P_r-1}
\end{eqnarray}
Here $X$ and $Y$ are some combinations of parameters of the BD and PD,
\begin{equation}
X = 1-p'+p'e^{-\lambda}\qquad{\rm and}\qquad Y = \frac{p'}{X}e^{-\lambda}, \label{XY-1}
\end{equation}
$S(N,r)$ are the Stirling numbers of the second kind defined in Section \ref{Sn} and $P_{BD}(r) =  \binom{K}{r}\left( \tilde{p}\right)^r \left( 1 - \tilde{p}\right)^{K-r}$ is the BD multiplicity distribution listed in Table \ref{Table_1} with
\begin{equation}
p \to \tilde{p} = \frac{p'}{1+p' e^{-\lambda}}. \label{tildep-1}
\end{equation}

The corresponding combinants $C^{\star}_N$ can be written in a similar way as:
\begin{eqnarray}
C^{\star}_N &=& K\frac{\lambda^N}{N!} \sum_{r=1}^N (-1)^{r-1}(r-1)! S(N,r) Y^r = \label{Cstar-BD-PD-1}\\
&=& \frac{\lambda^N}{N!} \sum_{r=1}^N C^{\star}_r(BD)\cdot r! e^{-r\lambda} S(N,r), \label{C=CBDfactor-1}
\end{eqnarray}
here $C^{\star}_r = (-1)^{r-1}\frac{K}{r}\left(\frac{\tilde{p}}{1-\tilde{p}}\right)^r$ are for the combinants for a BD multiplicity distribution listed in Table \ref{Table_1} with $p$ replaced by $\tilde{p}$.

The question that interests us the most now is how the combinants' oscillation period changes when switching from $P_{BD}$ to $P_{CD} = P_{BD}\circ P_{PD}$. To this end, we now return to Eq. (\ref{Cstar-BD-PD-1}) to note that the oscillations of $C^{\star}_j$ are given by the sum of the powers of a certain combinations of parameters of both distributions, $Y=\frac{p'e^{-\lambda}}{1 - p' + p'e^{-\lambda}}$ (as given by Eq. (\ref{XY-1})), the combinatorial part of which is
given by the Stirling numbers of the second kind. At this point, however, the possibility of further analytical examination of this sum ends. It can be further represented in various ways, e.g. using Eq. (\ref{WS2-1}) expressed via the polylogarithm
function $Li_j(x)$ as
\begin{equation}
C^{\star}_N =  K\frac{\lambda^N}{N!} \sum_{r=1}^N (-1)^{r-1}(r-1)! S(N,r) Y^r = K\frac{\lambda^N}{N!}\cdot (-1)^{N-1}Li_{1-N}\left( 1 - \frac{1}{Y}\right). \label{Cstar-BD-PD-1}
\end{equation}
In this case the general statement is that oscillations of $C^{\star}_N$ reflect oscilatory behaviour of this polylogarithmic function. The relation between the parameters $p'$ (from the BD) and $\lambda$ (from the PD) giving the same value of $Y$, therefore the same period $T$ of oscillation of $C^{\star}_N$, is
\begin{equation}
\lambda = \ln \left( \frac{1-Y}{Y}\right) + \ln \left( \frac{p'}{1-p'} \right). \label{plY}
\end{equation}
As for the dependence of period $T$ on parameter $Y$, it can be parameterized as
\begin{equation}
T = 2.4 - 1.73 \ln Y. \label{T(Y)}
\end{equation}

\subsection{Compound distribution $P_{CD} = P_{BD}\circ P_{NBD}$}
\label{CD=BDxNBD}

Using $F(z)$ and $G(z)$ for, respectively, BD and PD as given in Table \ref{Table_1} we have that the resulting multiplicity distribution $P(N)$
can be presented in one of the following forms (details of their derivation are presented in \ref{Derivation-2}):
\begin{eqnarray}
P(N) &=& \frac{p^N}{N!}\tilde{X}^K \cdot \sum_{r=0}^{min(N,K)} r!\binom{K}{r} B_{N,r}\left[ \left\{ k^{(N-r+1)}\right\}\right]\cdot \tilde{Y}^r =\label{BD-NBD-final-2}\\
&=& \frac{p^N}{N!}\left[ 1 + p'(1-p)^k\right]^K \cdot\nonumber\\
&&\hspace{15mm} \cdot \sum_{r=0}^{min(N,K)} P_{BD}(r) \cdot\left\{ r! (1-p)^{rk}B_{N,r}\left[ \left\{ k^{(N-r+1)}\right\}\right] \right\}, \label{Pr-BD*NBD-1}
\end{eqnarray}
where $k^{(m)} = \prod_{i=0}^{m-1}(k+i) = k(k+1)\cdots(k+m-1)$ is a rising factorial defined in Eq. (\ref{PS-rf}). Here $\tilde{X}$ and $\tilde{Y}$ are some combinations of parameters of the BD and NBD,
\begin{equation}
\tilde{X} = 1 - p' + p' (1 - p)^k\qquad{\rm and}\qquad \tilde{Y} = \frac{p'(1-p)^k}{\tilde{X}}, \label{XY-2}
\end{equation}
and $P_{BD}(r) =  \binom{K}{r}\left( \hat{p}\right)^r \left( 1 - \hat{p}\right)^{K-r}$ in the $r$-th component in the total of $P(N)$ has the form characteristic for the BD multiplicity distribution, as given in Table \ref{Table_1} with
\begin{equation}
p\to \hat{p} = \frac{p'}{1+p' (1-p)^k}. \label{tildepNBD-1}
\end{equation}

The corresponding combinants $C^{\star}_N$ can be written in a similar way as:
\begin{eqnarray}
C^{\star}_N &=& K \frac{p^N}{N!} \sum_{r=1}^N (-1)^{r-1} (r-1)! B_{N,r}\left[ \left\{ k^{(N-r+1)}\right\}\right]\cdot \tilde{Y}^r = \label{Cstar-BD-NBD-fin-1}\\
&=& \frac{p^N}{N!} \sum_{r=1}^N C^{\star}_r(BD)\cdot r! (1-p)^{rk} B_{N,r}\left[ \left\{ k^{(N-r+1)}\right\}\right], \label{C=CBDNBDfactor-1}
\end{eqnarray}
where the $C^{\star}_r(BD) = (-1)^{r-1} \frac{K}{r}\left(\frac{\hat{p}}{1-\hat{p}}\right)^r$ in the $r$-th component in the total of $C^{\star}_N$ has the form characteristic for the BD as given in Table \ref{Table_1} in which parameter $p'$ is replaced by $\hat{p}$.

As for the resultant oscillations of the combinants $C^{\star}_N$ given by Eq. (\ref{Cstar-BD-NBD-fin-1}), this time the parameter $\tilde{Y}$ depends on $3$ parameters: $p'$ (from the BD) and $p$ and $k$ (from the NBD). Therefore, the counterpart of Eq. (\ref{plY}) is now
\begin{equation}
(1-p)^k = \left( \frac{\tilde{Y}}{1 - \tilde{Y}}\right) + \left(\frac{1 - p'}{p'}\right), \label{ptildeY}
\end{equation}
with combinations of two parameters, $(1-p)^k$, replacing the action of the single parameter $\lambda$ before. Additionally, and this is the most important difference,  the combinatorial part represented by Bell polynomial now depends in a rather complicated way on the parameter $k$ of the NBD. As a result the nature of the oscillations is now much more complicated. Nevertheless, it is possible to extract some interesting features from these relations. For example, one can choose $\tilde{Y}$ and $k$ so as to have fixed period oscillations, $T$, and see what is the dependence of $\tilde{Y}$ on $k$. And so, for $T=12$ we have that $\tilde{Y} = 0.5\, k^{-1.46}$ whereas for $T=8$ we have $\tilde{Y} = 0.5\, k^{-0.7}$.

\section{Oscillations of combinants from the perspective of a simple toy model}
\label{Toy-model}

In order better to illustrate the mechanism of the formation of oscillations of combinators and understand what information is encoded in their period, we now refer to the model defined by the following generating function:
\begin{eqnarray}
G(z) &=& \sum_{i=0}^{\infty} A_i\cdot z^i\quad \stackrel{z \to 0}{\longrightarrow}\quad A_0, \label{gz-clust}\\
\frac{d^j G(z)}{d z^j} &=& \sum_{i=j}^{\infty} A_i\cdot  j!\binom{i}{j} z^{i-j}\quad \stackrel{z\to 0}{\Longrightarrow}\quad j! A_j. \label{gz-clust-der}
\end{eqnarray}
This model describes the situation where particles are produced in $i$-particle clusters, the distribution of which is given by a BD (described by parameters $p$ and $K$) with generating function $F[G(z)] =  \left[ p G(z) + 1 - p\right]^K$ and $L(z) = \ln \{ F[G(z)]\}$ for which
\begin{eqnarray}
\frac{d^n F}{d^n z} &=& n!\binom{K}{n}p^n \left[ pG(z) + 1 - p\right]^{K-n}\quad \stackrel{z \to 0}{\Longrightarrow}\nonumber\\
&& \stackrel{z \to 0}{\Longrightarrow} \quad n!\binom{K}{n}p^n (1 - p)^{K-n} = n! P_{BD}(n|K,p),\label{DFn-2a}\\
\frac{d^n L}{d z^n} &=& (-1)^{n-1} (n-1)!~K \left[ \frac{p}{pG(z) + 1 - p}\right]^n \quad \stackrel{z \to 0}{\Longrightarrow}\nonumber\\
&& \stackrel{z \to 0}{\Longrightarrow}\quad  (-1)^{n-1} (n-1)!~K \left( \frac{p}{1 - p}\right)^n  = n! C^{\star}_n(BD|K,p). \label{DL-na}
\end{eqnarray}
The coefficients $A_i$ are equal to the probabilities of the respective multiplicities, $A_j = P(j)$, where the $P(j)$ is the multiplicity distribution of the process considered here.} The resultant $P(N)$ and $C^{\star}_n$ for the CD composed with this BD (with parameters $K$ and $p$) and this kind of the hadronizing cluster hadronization process are\footnote{
Note that Eq. (\ref{gz-clust}) and Eqs. (\ref{BDclust-P(N-1)}), (\ref{BDclust-Cstar-1}) do indeed describe a quite general situation. In particular, for $A_j = P_{PD}(j)$ and for $A_j = P_{NBD}(j)$ we get the previously discussed results for, respectively, $P_{CD} = P_{BD}\circ P_{PD}$ (summarized in Eqs. (\ref{P_r}) and (\ref{C=CBDfactor})) and for $P_{CD}(N) = P_{BD}\circ P_{NBD}$ (summarised in Eqs. (\ref{Pr-BD*NBD}) and (\ref{C=CBDNBDfactor}))}.
\begin{eqnarray}
P(N) &=& \frac{1}{N!}\frac{d^N F[G(z)]}{d z^N}\bigg|_{z=0} = \frac{1}{N!}\sum_{r=1}^N r! P_{BD}(r)\cdot B_{N,r}\left( A_1, A_2,\dots,A_{N-r+1} \right),\label{BDclust-P(N-1)}\\
 C^{\star}_N &=& \frac{1}{N!}\frac{d^N L[G(z)]}{d z^N}\bigg|_{z=0} = \frac{1}{N!}\sum_{r=1}^N r! C^{\star}_r(BD)\cdot B_{N,r}\left( A_1, A_2,\dots,A_{N-r+1} \right).  \label{BDclust-Cstar-1}
\end{eqnarray}

We shall now use this model in a simplified version as a particular toy model analytically to calculate the two simplest situations:\\
$(i)$ Only $a$-particle clusters are produced, i.e., $i=a$, and
\begin{equation}
G(z) = z^a,\qquad \frac{d^j G(z)}{d z^j} = a(a-1)\cdots (a-j+1) z^{a-j}\quad \stackrel{z\to 0}{\Longrightarrow}\quad = j!\delta_{j,a}, \label{z-t0-a}
\end{equation}
$(ii)$ The particles are produced only in $a$-particle and $b$-particle clusters, i.e, $i=a,b$, and
\begin{equation}
G(z) = A_a z^a + A_b z^b,\qquad A_a + A_b = 1,\qquad \frac{d^j(G(z)}{d z^j}\bigg|_{z=0} = a! A_a \delta_{j,a} + b! A_b \delta_{j,b.}. \label{z-t0-a-and-b}
\end{equation}
In the first case
\begin{eqnarray}
P(N) &=& \frac{1}{N!}\frac{d^N F[G(z)]}{d z^N}\bigg|_{z=0} = \frac{1}{N!}\sum_{r=1}^N r! P_{BD}(r)\cdot B_{N,r}\left(0,\dots, x_a=1,\dots,0 \right) = \nonumber\\
&=& \frac{1}{N!}\sum_{r=1}^N \delta_{ar,N} \cdot P_{BD}(r)\cdot \frac{(ar)!}{(a!)^r} =\nonumber\\
&=&  \left\{
\begin{array}{rl}
P_{BD}\left(r=\frac{N}{a}\right)\cdot \frac{1}{(a!)^{\left(\frac{N}{a}\right)}}\quad & \text{for  $N$ being integer multiple}\quad a\\
0\qquad\qquad\qquad\qquad & \text{otherwise}
\end{array}
\right. ,  \label{X}
\end{eqnarray}
\begin{eqnarray}
 C^{\star}_N &=&\frac{1}{N!}\frac{d^N L[G(z)]}{d z^N}\bigg|_{z=0} = \frac{1}{N!}\sum_{r=1}^N r! C^{\star}_r(BD)\cdot B_{N,r}\left(0,\dots, x_a=1,\dots,0 \right) = \nonumber\\
 &=& \frac{1}{N!}\sum_{r=1}^N \delta_{ar,N} \cdot C^{\star}_r(BD)\cdot\frac{(ar)!}{(a!)^r} =\nonumber\\
&=&  \left\{
\begin{array}{rl}
C^{\star}_{r= \frac{N}{a}}(BD)\cdot \frac{1}{(a!)^{\left(\frac{N}{a}\right)}}\quad & \text{for  $N$ being integer multiple}\quad a\\
0\qquad\qquad\qquad\quad & \text{otherwise}
\end{array}
\right.  . \label{Y}
\end{eqnarray}
Here we have taken advantage of the fact that the condition (\ref{Def-1a}) in the definition (\ref{Def-1}) of the Bell polynomials requires that if only a single $x_j\neq 0$ then $k_j=k$ and $j k_j = j k =n$, which immediately results in Eq. (\ref{Def-SS1-a}),
which in our case has the form
\begin{equation}
B_{n,k}\left(0,\dots,x_j,\dots, 0\right) = 0,~{\rm except~when}~ n=jk~ {\rm and}~B_{jk,k}=\frac{(jk)!}{k!}\left(\frac{x_j}{j!}\right)^k. \label{single-x}
\end{equation}
To get the proper $P(0)$ too we have to include the case with $r=N=0$.  Because $0!=1$ therefore $P(0) = P_{BD}(0) = (1-p)^K$, as it should be. The expression for combinants remains intact.

This example shows that while for $a=1$ we just reproduce the initial multiplicity distribution and combinants, i.e., $P(N)=P_{BD}(N)$ and $C^{\star}_N = C^{\star}_{BD}(N)$ with period equal $2$, for $a>1$ the period of the oscillations of the combinants changes to $2a$. Therefore, if we observe oscillations with the period $2a$ in the data, it can be assumed that we are dealing with the production of $a$-partial clusters.

In the second case, we have
\begin{eqnarray}
&&\!\!\!\!\!\!\! P(N)\! =\! \frac{1}{N!}\frac{d^N F[G(z)]}{d z^N}\bigg|_{z=0} = \frac{1}{N!}\sum_{r=1}^N r! P_{BD}(r) B_{N,r}\left(0,\dots,A_a,\dots,A_b, \dots,0 \right), \label{BDclust-P(N)-2a1}\\
&&\!\!\!\! C^{\star}_N\! =\! \frac{1}{N!}\frac{d^N L[G(z)]}{d z^N}\bigg|_{z=0} = \frac{1}{N!}\sum_{r=1}^N r! C^{\star}_r(BD) B_{N,r}\left(0,\dots,A_a,\dots, A_b,\dots,0 \right), \label{BDclust-Cstar-2a1}
\end{eqnarray}
where
\begin{equation}
B_{N,r}\left(0,\dots,A_a,\dots,A_b,\dots,0\right) = \sum_{k_a,k_b}\frac{N!}{k_a! k_b!} \left(\frac{A_a}{a!}\right)^{k_a}\left(\frac{A_b}{b!}\right)^{k_b},  \label{BNr-ka-kb}
\end{equation}
and where, for given $N$, $a$ and $b$, the summation is over all non-negative integers $k_{a,b}$ satisfying the conditions
\begin{equation}
k_a + k_b = r\qquad {\rm and}\qquad ak_a + bk_b =N.  \label{cond-a2-1}
\end{equation}
Note that the second condition tells us that $P(N)$ and $C^{\star}_N$ are nonzero only for multiplicities $N$ equal to the sum of multiple $a$ and multiple $b$. In conjunction with the first condition, this formally allows us, for example, to assume that the
\begin{equation}
k_a = \frac{br - N}{b-a} \label{k_a}
\end{equation}
is given and the summation is only over $k_b$. However, we must also take into account that both $k_{a,b}$ are non-negative integers, which can be written using the Kronecker delta function, $\delta_{(b-a)k_a,br - N}$. For the given values of $a$ and $b$, this imposes limits on the allowable values of $N$ and $r$. If they are not fulfilled, these $(N, r)$ will not appear in Eq. (\ref{BDclust-C(N)-4}) below, significantly influencing the nature of the oscillations. So we can write the final result analytically as
\begin{eqnarray}
P(N) &=&  \sum_{r=1}^N P_{BD}(r)\left( \frac{A_{b}}{b!}\right)^r \sum_{k_a=0}^r \binom{r}{k_a} \left( \frac{A_a}{A_b}\frac{b!}{a!}\right)^{k_a} \cdot \delta_{(b-a)k_a,br - N}. \label{BDclust-P(N)-4}
\end{eqnarray}
\begin{eqnarray}
 C^{\star}_N &=& \sum_{r=1}^N C^{\star}_r(BD)\left( \frac{A_{b}}{b!}\right)^r \sum_{k_a=0}^r \binom{r}{k_a} \left( \frac{A_a}{A_b}\frac{b!}{a!}\right)^{k_a} \cdot \delta_{(b-a)k_a,br - N}. \label{BDclust-C(N)-4}
\end{eqnarray}
In this case the oscillations of $C_N^{\star}$ will have, depending on the values of parameters $a$ and $b$ used, a more or less complex and non-uniform character. This example basically exhausts the list of variants of toy models that can be calculated analytically.

\section{Examples of other possible types of combined multiplicity distributions}
\label{OtherCMD}

We shall now present, for the sake of completeness, two examples of distributions leading to oscillating combinants, both being some complex distributions but not the CDs discussed above: a weighted sum of multiplicity distributions as a quasi-BD and the sum of the multiplicities from distributions of different types.

\subsection{Weighted sum of multiplicity distributions as a quasi-BD}
\label{QuasiBD}

The first example comes from Refs. \cite{Ours1,Zborovsky} showing that oscillations of combinants can also occur in multiplicity distributions $P(N)$ expressed as a weighted sum of a number of distributions $P_i(N)$, none of which individually leads to such oscillations. In particular, Ref. \cite{Zborovsky} shows that the oscillations observed in the LHC data can be very well described by an appropriately selected weighted sum of three NBDs. In both works the combinants were deduced using the recurrence relation given by Eq. (\ref{rCj}). Using the approach based on generating functions, it can be shown that the oscillations appear because, in fact, the superposition of a number of distributions $P_i(N)$ leads to the same expressions for combinants as we get from the previously discussed compound distributions, which were based on the BD.

To show this, let us assume that $P(N)$ is a weighted sum of $n$ components $P_i(N)$ ($\{ p_i\}$ denote summarily all parameters of the consecutive components),
\begin{equation}
P(N) = \sum_{i=1}^n w_i P_i\left(N,\langle N_i\rangle,\{ p_i\}\right),\qquad \sum_{i=1}^n w_i = 1, \label{Sum-1}
\end{equation}
with corresponding generating function and combinants,
\begin{eqnarray}
H(z) &=& \sum_{i=1}^n H_i(z) = \sum_{i=1}^n w_i G_i(z)\qquad {\rm and}\qquad C^{\star}_N = \frac{1}{N!}\frac{d^N \ln [H(z)]}{d z^N} \bigg|_{z=0}, \label{Sum-3}\\
\frac{d^N \ln[H(z)]}{d z^N} &=& \sum_{r=1}^N \left[ \frac{d^r \ln [H(z)]}{d H^r}\right]\cdot
B_{N,r}\left[ H^{(1)}(z),H^{(2)}(z),\dots,H^{(N-r+1)}(z)\right] =\nonumber\\
&=& \sum_{r=1}^N \left\{(-1)^{r-1}(r-1)! [H(z)]^{-r} \right\}\cdot \mathbb{R}, \label{Sum-6}\\
{\rm where}\quad \mathbb{R} &=&  \sum_{\left\{r_1,\dots,r_N \right\}}
 \left( \frac{N!}{r_1!\cdots r_N!} \right) \prod_{m=1}^{N-r+1} \left[ \frac{H^{(m)}(z)}{m!}\right]^{r_m},\quad
H^{(i)}(z) = \frac{d^i H(z)}{d z^i}, \nonumber
\end{eqnarray}
and summation is over all $\left\{ r_{i=1,\dots,N+r-1}\right\}$ such that $\sum_{m=1}^N = m r_m = N$ and  $\sum_{m=1}^N r_m= r$.
Now, note that in Eq. (\ref{Sum-6}) the factor $(-1)^r$ appeared coming from the $r-$th derivative of the logarithm of the generating function $H(z)$, which is responsible for the oscillations of $C^{\star}_N$ ($H^{(i)}(0)=H^{(i)}(z=0)$):
\begin{equation}
C^{\star}_N = \sum_{r=1}^N (-1)^{r-1}\frac{(r-1)!}{[H(0)]^r} B_{N,r}\left[ H^{(1)}(0), H^{(2)}(0),\dots, H^{(N-r+1)}(0)\right]. \label{Sum-8}
\end{equation}
Note also that if all $n$ components of $P(N)$ are identical NBD then the resultant $P(N)$ is a NBD with $H(z) = G(z)$ as given by Eq. (\ref{G-NBD}) and with $C^{\star}_N$ from Eq. (\ref{Sum-3}) coinciding with the respective $C^{\star}_N$ for a NBD and presented in Table \ref{Table_1}, which do not oscillate. Otherwise, there will be oscillations that will depend on how much the parameters $\{ p_i\}$ differ from each other and in what proportions $\{ w_i\}$ individual NBDs appear. On the example of Ref. \cite{Zborovsky}, we can closely trace the process of formation of these oscillations and the changes in their character when we add a new component.

\subsection{The sum of the multiplicities from distributions of different types}
\label{SdiffP}

The second example uses a distribution of the sum of multiplicities from distributions of different types, for example BD and NBD as used by us in Refs. \cite{Ours3,Ours-EPJA}. In this case, the corresponding generating function is the product of the generating functions of the individual distributions. In the above example
\begin{equation}
G(z)=G_{BD}(z)G_{NBD}(z), \label{GBDNBD}
\end{equation}
the resulting multiplicity distribution is the convolution of the corresponding multiplicity distributions,
\begin{equation}
P(N) = \sum_{i=0}^{min\left\{ N,K\right\}} P_{BD}(i)P_{NBD}(N-i), \label{PbdPnbd}
\end{equation}
the observed multiplicity is sum of the multiplicities from both components,
\begin{equation}
N = N_{BD} + N_{NBD}, \label{NN}
\end{equation}
and the respective modified combinants are the weighted sums of the combinants for both components,
\begin{equation}
\langle N\rangle C_j = \left< N_{BD}\right>C_j^{(BD)} + \left< N_{NBD}\right> C_j^{(NBD)}. \label{Cjbdnbd}
\end{equation}
This approach allows us to explain the experimental situation mentioned in Section \ref{Survey} when we observe the oscillations of $C_j$ with period $2$ but the condition (\ref{TwoMoments}) is not fulfilled, i.e. $Var(N)/\langle N\rangle > 1$.
In the approach proposed here, this would mean the domination of the BD component in $P(N)$, the second component would only affect the oscillation amplitude. Note that this distinguishes this type of description from the CDs discussed above, which always lead to oscillations with period greater than $2$.

\section{Interconnections of multiplicity distributions and combinants using Bell polynomials}
\label{Inter}

For the sake of completeness of our presentation, we now briefly present how the use of Bell polynomials allows for the formal expression of multiplicity distributions $P(N)$ by their combinants $C^{\star}_N$ (and vice versa). Following Refs. \cite{Combinants-2,VVP} and using the generating function approach one can, in a sense, solve recurrences (\ref{rr}) and (\ref{rCj}) and obtain that:
\begin{eqnarray}
P(N) &=& P(0) \frac{1}{N!} \sum_{j=0}^{\infty} B_{N,j}\left[1! C_1^{\star},\dots,(N-j+1)! C_{N-j+1}^{\star}\right], \label{PfromC}\\
C_N^{\star} &=& \frac{1}{N!}\sum_{j=1}^N (-1)^{j+1} (j-1)!\cdot\nonumber\\
 && \qquad \cdot B_{N,j} \left\{1! \left[ \frac{P(1)}{P(0)}\right],\dots,(N-j+1)! \left[ \frac{P(N-j+1)}{P(0)}\right]\right\}.\label{CfromP}
\end{eqnarray}
We shall illustrate this by determining these values for the distributions BD, PD and NBD considered here.

For the  BD we have $N! C^{\star}_N = (-1)^{N-1}(N-1)! K\left(\frac{p}{1-p}\right)^N$ and $N! \frac{P(N)}{P(0)} = (K)_n \left( \frac{p}{1-p}\right)^N$, where $(K)_N = K(K-1)(N-2)\cdots(K-N+1) = \frac{\Gamma(K+1)}{\Gamma(K-N+1)}$ are falling factorials defined in Eq. (\ref{PS-ff}). Therefore
\begin{eqnarray}
    P(N) &=&  (1 - p)^K \frac{1}{N!}\cdot (-1)^N\left(\frac{p}{1-p}\right)^N \sum_{j=0}^N (-1)^j K^j B_{N,j}[0!,1!,\dots,(N-j)!] = \nonumber\\
    &=&  p^N (1 - p)^{K-N} \frac{1}{N!}\cdot \sum_{j=0}^N K^j s(N,j) =  \binom{K}{N}p^N (1 - p)^{K-N},            \label{BD-1}
\end{eqnarray}
where we used the following formulas in turn: Eq. (\ref{Def-3}), Eq. (\ref{Def-5}) and Eq. (\ref{def-ssnfk})
  ($s(N,j)$ are signed Stirling numbers of the first kind defined in Section \ref{Sn}). Correspondingly
    \begin{equation}
    C_N^{\star} = \frac{1}{N!}\left( \frac{p}{1-p}\right)^N \sum_{j=1}^N (-1)^{j+1} (j-1)! B_{N,j}\left[(K)_1,\dots,(K)_{N-j+1}\right].\label{BD-2}
    \end{equation}
However, this time we did not  find an analytic expression for the Bell polynomial used here, so we checked by direct calculation that  we reproduce using Eq. (\ref{BD-2}) the combinants $C_N = \frac{N+1}{\langle N\rangle}C^{\star}_{N+1}$ from Table \ref{Table_1}. Because $C^{\star}_N = (-1)^{N-1}\frac{K}{N}\left( \frac{p}{1-p}\right)^N$ therefore we consider that
\begin{equation}
\sum_{j=1}^N (-1)^{j-1} (j-1)! B_{N,j}\left[(K)_1,\dots,(K)_{N-j+1}\right] = (-1)^{N-1}(N-1)!\, K,
\label{NewResBP-1}
\end{equation}
is a new rule for Bell polynomials (to be added to the list provided in Section \ref{BP}).

For the PD we have $N! C^{\star}_N = N!\,\lambda \delta_{N,1}$ and $N! \frac{P(N)}{P(0)} = \lambda^N$. Therefore using Eq. (\ref{Def-SS1}) we have
\begin{equation}
P(N) = e^{-\lambda} \frac{1}{N!}\sum_{j=0}^N B_{N,j}(\lambda,0,\dots,0)= e^{-\lambda} \frac{1}{N!} \sum_{j=0}^N \lambda^j \delta_{N,j} = \frac{\lambda^N}{N!}e^{-\lambda}. \label{PD-1}
\end{equation}
Correspondingly, using sequentially Eqs. (\ref{Def-4}), (\ref{S1}), (\ref{special-1}) and (\ref{WS2-2}) we have
\begin{eqnarray}
C^{\star}_N = \frac{\lambda^N}{N!}\sum_{j=1}^N (-1)^{j-1}(j-1)!B_{N,j}(1,\dots,1) &=& \frac{\lambda^N}{N!} \sum_{r=1}^N (-1)^{r-1} (r-1)! S(N,r) = \nonumber\\
&=& \lambda \delta_{N,1}. \label{PD-2}
\end{eqnarray}

For the NBD we have that $N! C^{\star}_N = (N-1)! kp^N$ and $N! \left[ \frac{P(N)}{P(0)}\right] =  k^{(N)}\cdot p^N$,
where $k^{(N)} = k(k+1)(k+2)\cdots(k + N -1) = \frac{\Gamma(k+N)}{\Gamma(k)}$ are rising factorials defined in Eq. (\ref{PS-rf}). Therefore
\begin{eqnarray}
P(N) &=& (1 - p)^k \frac{1}{N!}\sum_{j=0}^N B_{N,j}[kp, kp^2,\dots,(N-j)!kp^{N-j}] = \nonumber\\
&=& (1 - p)^k \frac{p^N}{N!}\cdot \sum_{j=0}^N k^j B_{N,j}[0!,1!,\dots,(N-j)!] = \nonumber\\
&=& (1-p)^k \frac{p^N}{N!}\cdot \sum_{j=0}^N k^j |s(N,j)| = \binom{N+k-1}{N} p^N (1-p)^k. \label{NBD-1}
\end{eqnarray}
Correspondingly, using in turn: Eq. (\ref{Def-3}), Eq. (\ref{Def-5}) and Eq. (\ref{def-usstfk}) ($|s(N,j)|$ are unsigned Stirling numbers of the first kind defined in Section \ref{Sn}), we have
\begin{eqnarray}
C_N^{\star} &=&  \frac{p^N}{N!}\sum_{j=1}^N (-1)^{j+1} (j-1)!B_{N,j}\left[k^{(1)},\dots,k^{(N-j+1)}\right].\label{NBD-2}
\end{eqnarray}
The same remark as for Eq. (\ref{BD-2}) also applies here and because $C^{\star}_N = \frac{1}{N}kp^N$  we consider that
\begin{equation}
\sum_{j=1}^N (-1)^{j+1} (j-1)! B_{N,j}\left[k^{(1)},\dots,k^{(N-j+1)}\right] = (N-1)!\,k, \label{NewResBP-2}
\end{equation}
is another new rule for Bell polynomials (to be added to the list provided in Section \ref{BP}).

\section{Summary and discussion}
\label{SumDisc}

In conclusion, we have presented a possible view of the experimentally observed multiplicity distributions, $P(N)$, which takes into account the, perhaps still underestimated, role of the combinants $C^{\star}_j$ obtained from these distributions, in particular as regards information about the dynamics of the multiparticle production processes hidden in their oscillations. Such oscillations appear in virtually all observed reactions. The only condition is, as shown in \cite{Zborovsky,Ours2}, sufficiently large statistics and good control over systematic errors. The better to recognize the role of the individual parameters describing the distributions used, we tried to stay at the level of analytical calculations.

\begin{figure}[b]
\vspace{-10mm}
\centering
\includegraphics[scale=0.55]{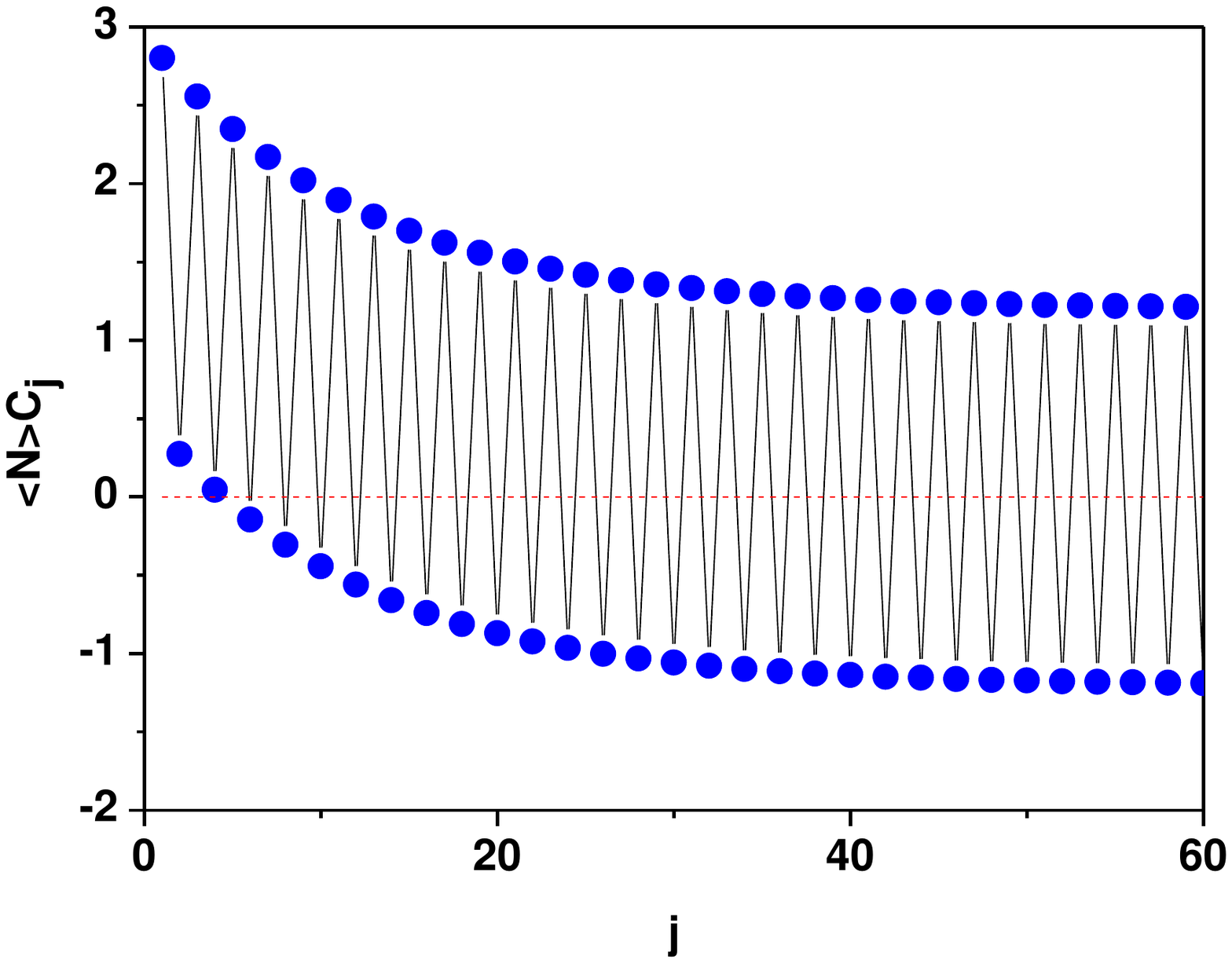}
\vspace{-75mm}
\caption{The  modified combinants $C_j$  from the generating function given by the product  $G(z) = G_{BD}(z)G_{NBD}(z)$ with moments of multiplicity distributions:  $\langle N_{BD} \rangle = 0.6$ and $Var(N_{BD}) = 0.3$ for the BD and $\langle N_{NBD}\rangle =20$ and $Var(N_{NBD}) = 250$ for the NBD.  \label{Fig_BD+NBD}}
\end{figure}
The most important conclusion from our analysis is that in order to describe both the measured $P(N)$ and their combinants, it is necessary to use phenomenological multiplicity distributions that are some compound distributions based on the BD. We arrive at this conclusion by noting that combinants corresponding to 'pure' BD, PD or NBD distributions do not provide more information than we can get from the multiplicity distributions $P(N)$. As shown in Eq. (\ref{TwoMoments}) they are fully determined by $\langle N\rangle$ and $Var(N)$ for the appropriate distributions $P(N)$. However, while the BD always gives oscillations with period $2$, experimentally we usually observe oscillations with greater periods. We interpret this fact as the already mentioned presence in $P(N)$ of some additional source of fluctuations resulting in a compound distribution. This is where the key role of combinants becomes apparent, because they provide us with information that is otherwise unavailable from the observed distributions $P(N)$. It should be emphasized that staying at the level of analytical calculations made it possible to obtain for the considered examples of combined distributions transparent formulas for multiplicity distributions and the corresponding combinants. They show exactly which combinations of parameters in the distributions included in a given CD are essential and what are the combinatorial elements (described herein by Bell polynomials) found in these CDs. These results may provide a better understanding of the numerical algorithms commonly used to fit experimental data, and in particular they should make it significantly easier to find the best combinations of parameters used.

\begin{figure}[t]
\centering
\includegraphics[scale=0.55]{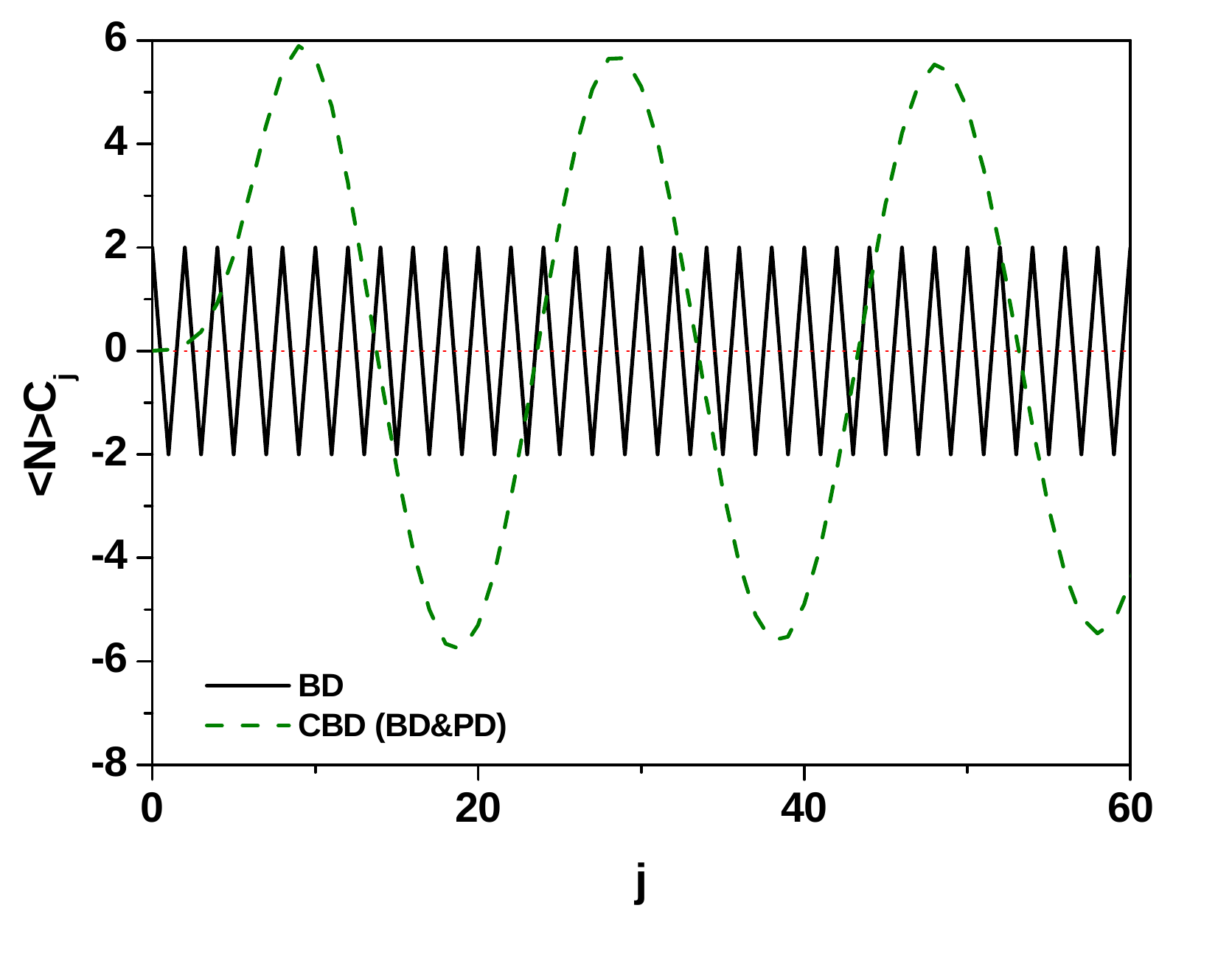}
\vspace{-5mm}
\caption{Modified combinants $C_j$ for a pure BD with $\langle N_{BD}\rangle = 1.5$ and $Var(N_{BD}) = 0.75$ compared with $C_j$
for a compound distribution $P_{BD}\circ P_{PD}$ with parameters $\langle N_{BD}\rangle = 1.8$ and $Var(N_{BD}) =0.72$ for BD and
$\lambda = 10$ for PD.  \label{Fig_BD&NBD}}
\end{figure}
To illustrate the key role of combinants mentioned above let us take the example of the fluctuations shown in Fig. \ref{Fig_BD+NBD}. We can see here oscillations with period $2$, as in the case of the BD, but with a varying amplitude unusual for a BD. It turns out, however, that if we take the NBD (which gives monotonically decreasing, non-oscillating $ C_j $) and add to it some particles from a BD (following the recipe from Section \ref{SdiffP})), we get just such a picture of $ C_j $. This therefore means that if we observe $C_j$ oscillations with period $2$ (like in a "pure BD") but generally looking like Fig. \ref{Fig_BD+NBD} then we can consider it as a signal that we are dealing with some distribution $P(N)$ where one of its components is a small admixture of particles from a BD dominating the shape of the amplitudes of such $C_j$. If we observe oscillations with a period greater than $2$, the situation is more interesting. In this case, most often we deal with a CD based on a BD, $H (z) = F_{BD}[G (z)]$, in which the oscillation period is related to the mean value of the distribution given by the generating function $G(z)$.
An example of such a case is presented in Fig. \ref{Fig_BD&NBD} in which we observe for the Binomial Compound Distribution $P_{BD}\circ P_{PD}$ oscillations with period $T=19$ as given by Eq.(\ref{T(Y)}). Thus, both of these examples clearly show how, by analyzing the behavior of $C_j$, one can extract information about the multiplicity distributions that make up the observed distribution $P(N)$.

\begin{figure}[b]
\vspace{-15mm}
\centering
\includegraphics[scale=0.5]{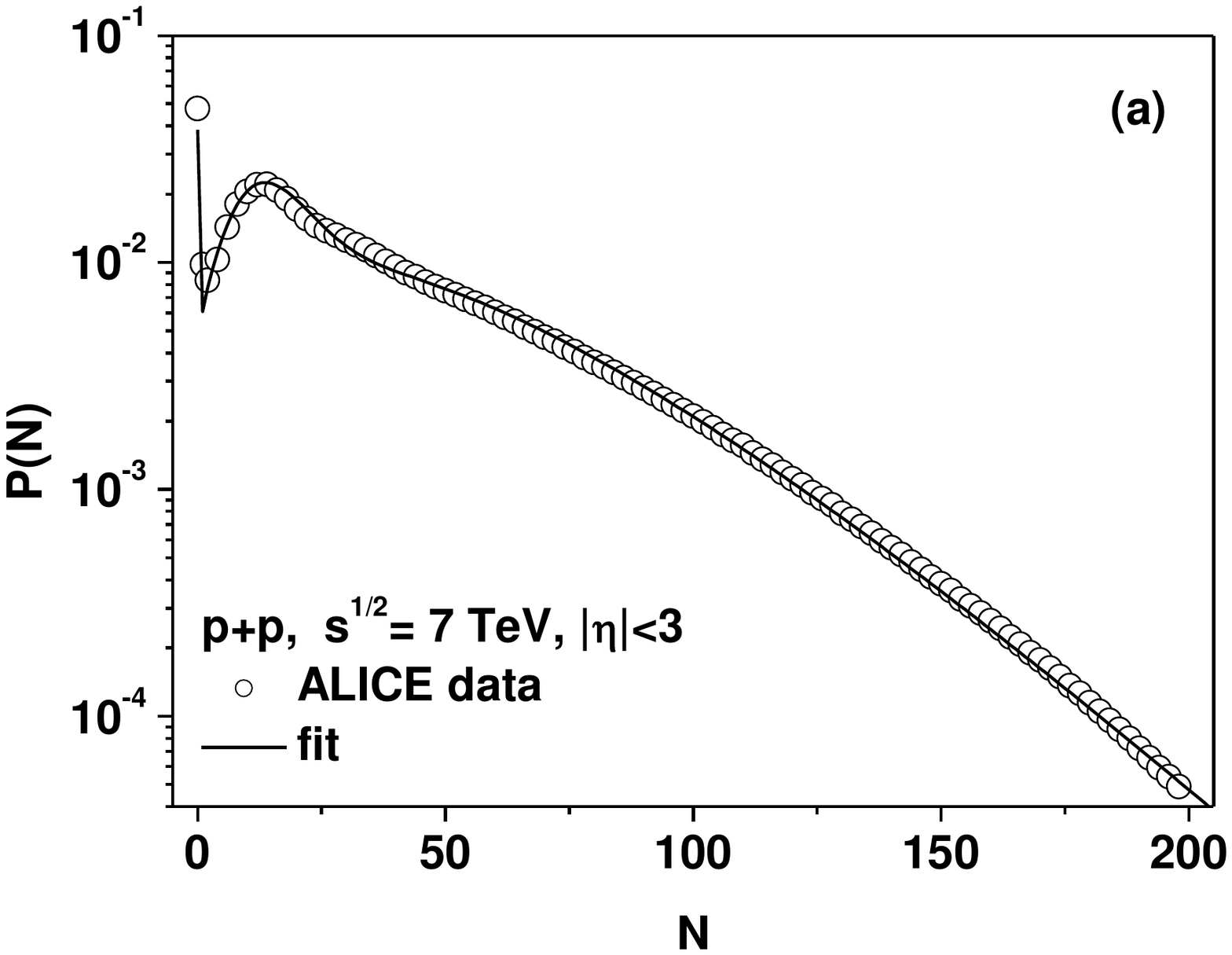}\\
\vspace{-7.5cm}
\includegraphics[scale=0.5]{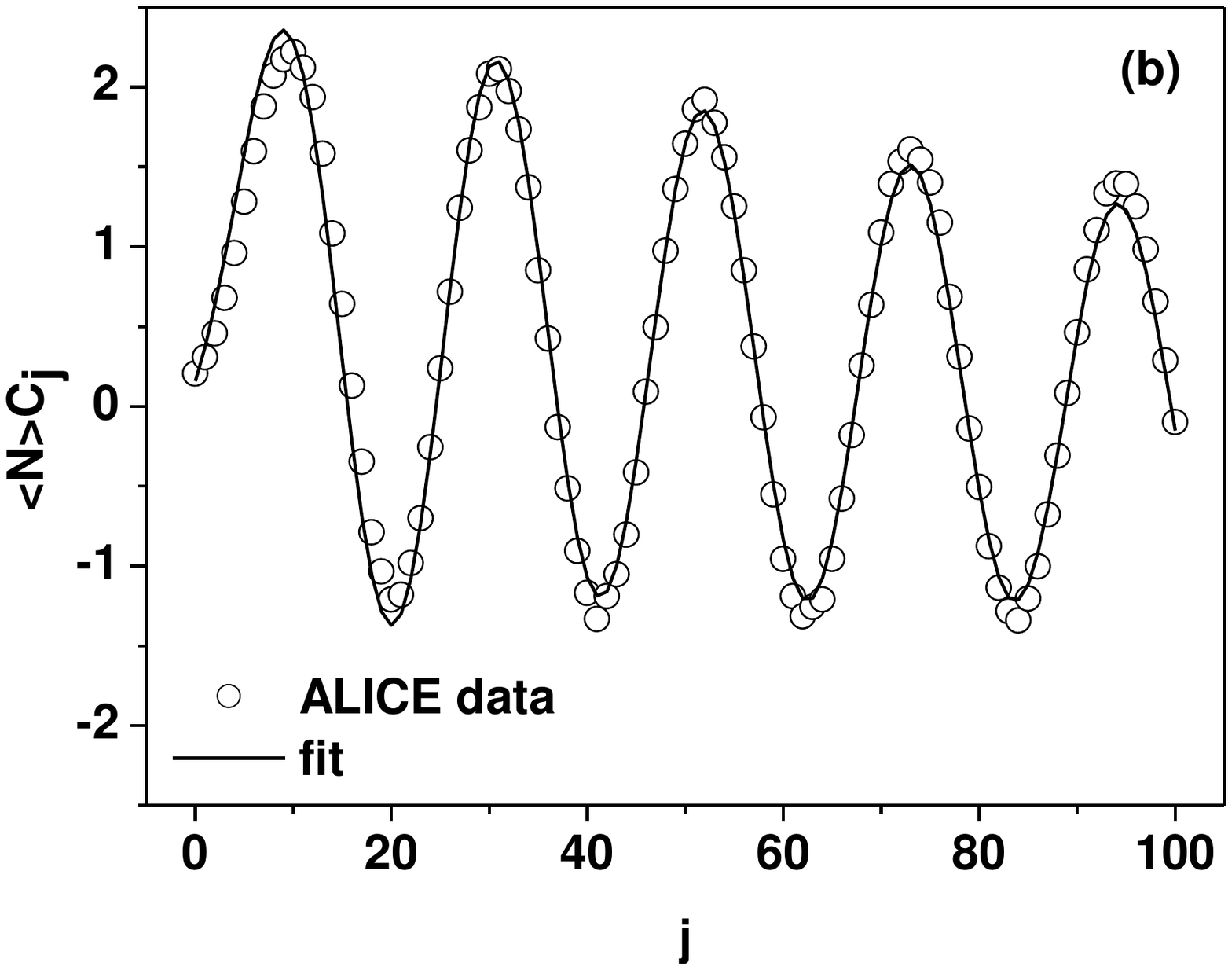}
\vspace{-6.7cm}
\caption{$(a)$ Multiplicity distributions $P(N)$ measured by ALICE \cite{ALICE} in $pp$ collisions at $\sqrt{s}=7$ TeV and pseudorapidity
window $|\eta|< 3$ (for clarity, every third point is shown only). $(b)$ The corresponding modified combinants $C_j$. Data are fitted using sum of two binomial compound distributions (BD\&NBD) given by Eqs. (\ref{BD-NBD-final-2}) and (\ref{Sum-1}) with parameters: $K_1 = K_2 = 3$,  $p_1 = 0.9$, $p_2 = 0.645$, $k_1 = 2.8$, $k_2 = 1.34$, $p_1'=0.67$, $p_2'=0.946$,  $w_1 = 0.24$ and $w_2 = 0.76$.  \label{Fig_Example}}
\end{figure}

The formalism of counting distributions presented here is meant to be used in high-energy particle physics. So far, the compound distributions CD presented here (one way or another based on BD) have been used to describe the phenomenon of the oscillatory behavior of combinants deduced from the multiplicity distributions measured in high energy $pp$ collisions at CMS and ALICE experiment at LHC CERN \cite{RWW-S,Ours2,Ours3,Ours-Odessa,Ours-EPJA}, in $e+e-$ annihilation measured in ALEPH experiment at LEP at CERN \cite{Ours3,Ours-Odessa,Ours-EPJA} and in $p\bar{p}$ collisions from the UA5 experiment at Super Proton-Antiproton Synchrotron at CERN \cite{Ours2,Ours3,Ours-Odessa,Ours-EPJA}. In paper \cite{Zborovsky} discussing $pp$ results obtained by ATLAS experiment at LHC a weighted sum of three NBDs has been used, which can be regarded, as we presented in Section \ref{QuasiBD}, as a variant of CD type of distribution based on BD.
Here we would like to highlight the following fact. It turns out that in practically all works using CD's, even single such distribution allows for adjustments with data that are basically very good. However, when one looks at the results obtained in more detail, there is usually still some room for improvement and the best fit to the data is obtained by taking the weighted sum of two such CDs \cite{Ours2}, only then do we have sufficient flexibility to fine-tune the data. As example we show in Fig. \ref{Fig_Example} description of the multiplicity distribution measured in ALICE experiment \cite{ALICE} and the corresponding modified combinants obtained from them.

On the other hand, the apparent necessity of using such weighted sums od two CDs could also signal that there are two independent, parallel mechanisms of particle production in action \cite{Ours3}.  However, as in the approach outlined in Section \ref{SdiffP}, we can also do otherwise, namely we can replace the previous weighted sum of CD's with the sum of the multiplicities from different distributions, one of which is the BD-based CD. It would still have oscillating $C_j$ but now with controlled periods and amplitudes, and it would be responsible for the observed oscillatory nature of $C_j$. This approach ensures the flexibility mentioned above, but with only one distribution. In fact, this idea has been successfully tested in \cite{Ours-EPJA} and seems to be a very promising tool for data analysis (despite an obvious drawback from the point of view of this work, that it can no longer be written analytically).

\section*{Acknowledgments}

This research was supported in part by  GW grants UMO-2016/22/M/ST2/00176 and DIR/WK/2016/2018/17-1. We thank Dr Nicholas Keeley for reading the manuscript.

\appendix

\section{Additional information}
\label{AddInf}

\subsection{Bell polynomials}
\label{BP}

Bell polynomials are defined as \cite{Comtet,BP,BN-1,BN-2}:
\begin{equation}
B_{n,k}\left(x_1,x_2,\dots,x_{n-k+1}\right) = \sum_{\{k_j\}} \left[\frac{n!}{k_{1}!\cdot k_{2}!\cdots  k_{n} !} \right]\prod_{j=1}^{n-k+1}\left( \frac{x_j}{j!} \right)^{k_j} \label{Def-1}
\end{equation}
where the summation is over all integers $k_1,k_2,\dots,k_n \ge 0$ such that
\begin{equation}
k_1 + 2k_2 + \dots + n k_n = n\quad {\rm and} \quad k_1 + k_2 +\cdots + k_n = k.
\label{Def-1a}
\end{equation}
Here is a list of some of their useful properties \cite{Comtet,Snfk,Lah,FengQi}:
\begin{eqnarray}
&&\left[ \frac{n!}{k_1! k_2 ! \cdots k_n!} \right] = S(n,k), \label{Def-0}\\
&& B_{n,k}\left(ab x_1,ab^2 x_2,\dots,ab^{n-k+1} x_{n-k+1}\right) = a^k b^n B_{n,k}\left( x_1,x_2,\dots,x_{n-k+1}\right), \label{Def-3}\\
&& B_{n,k}\underbrace{(1,1,\dots,1)}_\text{\tiny{$n-k+1$~-~times}} = S(n,k), \label{Def-4} \\
&& B_{n,n}\left( x_1,\dots,x_n \right) = x_n, \label{Def-S1}\\
&& B_{n,k}(1,0,\dots,0) = \binom{0}{n-k} = \delta_{n,k}, \label{Def-SS1}\\
&& B_{n,k}\left(0,\dots,x_j,\dots, 0\right) = 0 \qquad {\rm except~when}\qquad B_{jk,k}=\frac{(jk)!}{k!(j!)^k}\left(x_j\right)^k, \label{Def-SS1-a}\\
&& B_{n,k} [0!,1!,...,(n-k)!] = |s(n,k)| = (-1)^{n-k}s(n,k), \label{Def-5}
\end{eqnarray}
Here $S(n,k)$ are the Stirling numbers of the second kind,  $|s(N,r)|$ are the so-called  unsigned Stirling numbers of the first kind (cf. \cite{Comtet,Snfk}), (\ref{Def-SS1-a}) is from \cite{Comtet}, p. 136. The more involved relations are \cite{FengQi}:
\begin{eqnarray}
&& B_{n,k}\left[(x)_1,(x)_2,\dots,(x)_{n-k+1} \right] = \frac{(-1)^k}{k!}\sum_{l=0}^k (-1)^l \binom{k}{l}(x l)_n, \label{Def-7}\\
&&B_{n,k}\left[x^{(1)},x^{(2)},\dots,x^{(n-k+1)} \right] = \frac{(-1)^k}{k!}\sum_{l=0}^k (-1)^l \binom{k}{l}(xl)^{(n)}, \label{Def-8}
\end{eqnarray}
where $(x)_i$ and $x^{(i})$ are, respectivqely, falling and rising factorials defined in Section \ref{PS} by Eqs. (\ref{PS-ff}) and (\ref{PS-rf}).

\subsection{Stirling numbers of the first kind, $s(n,k)$ and of the second kind, $S(n,k)$}
\label{Sn}

The signed Stirling numbers of the first kind, $s(n,k)$ (cf. \cite{Comtet,Snfk}), are generated by the falling factorial:
\begin{equation}
(x)_n = \prod_{k=0}^{n-1}(x - k) = \sum_{k=0}^n s(n,k) x^k = \frac{\Gamma(x+1)}{\Gamma(x-n+1)} = n!\binom{x}{n},\qquad (x)_0=1.\label{def-ssnfk}
\end{equation}
The unsigned Stirling numbers of the first kind, $|s(n,k)|$, are defined by the rising factorial:
\begin{equation}
x^{(n)} = \prod_{k=0}^{n-1}(x+k) = \sum_{k=0}^n|s(n,k)| x^k = \frac{\Gamma(x+n)}{\Gamma(x)} = n!\binom{x+n-1}{n},\qquad x^{(0)}=1. \label{def-usstfk}
\end{equation}
Some properties of the Stirling numbers of the first kind:
\begin{eqnarray}
s(n,k) &=& (-1)^{n-k}|s(n,k)|,\qquad \quad s(n,1) = (-1)^{n-1}(n-1)! \label{rel-suns}\\
s(n,n) &=& 1,\qquad s(1,k) = \delta_{1,k},\qquad s(n,0) = 0,\qquad s(0,0)=1, \label{snfk-prop1}
\end{eqnarray}

Stirling numbers of the second kind, $S(n,k)$, count the set partitions of $[n]= {1,2,3,\dots,n}$ which consists of exactly $k$ subsets of parts \cite{Gould}. By definition $S(n,k)=0$ if $k=0$ or $k>n$. For technical reasons one defines $S(0,0)=1$. They are defined by the so called Euler formula (cf. Eq. (9.21) in \cite{Gould}):
\begin{equation}
S(n,k) = \left\{ \begin{array}{c}
n\\
k
\end{array}\right\} = \frac{(-1)^k}{k!} \sum_{i=0}^{k} (-1)^i \binom{k}{i}i^n = \frac{1}{k!}\sum_{i=0}^{k}(-1)^i\binom{k}{i}(k-i)^n, \label{S1}
\end{equation}
or by the recurrence relation \cite{Gould,WolframS2}:
\begin{eqnarray}
S(n+1,k) &=& k S(n,k) + S(n,k-1), \qquad 0 \leq k - 1 \leq n; \label{S2}\\
S(n,k) &=& \sum_{m=k}^n k^{m-m} S(m-1,k-1). \label{S2a}
\end{eqnarray}
From Eq. (\ref{S1}) one gets that \cite{BOY}
\begin{equation}
(-1)^r r!S(N,r) = \sum_{l=0}^r \binom{r}{l}(-1)^l l^N. \label{S1-a}
\end{equation}
Some properties of the Stirling numbers of the second kind:
\begin{eqnarray}
&&\hspace{-5mm} S(n,1)= S(n,n)=1,\quad S(n,2) = 2^{n-1} - 1,\quad S(n,n-1) = n(n-1)/2, \label{special-1}\\
&&\hspace{-5mm} S(0,0) = 1,\qquad\qquad S(n,0) = 0\qquad S(0,n) = 0, \label{special-2}
\end{eqnarray}
Using Eq. (\ref{S2}) one can calculate all numerical values for $S(n,k)$ (see examples in \cite{SNtable}). Some other useful properties of $S(n,k)$ are \cite{VCN} (($Li_{n}(z)$ denotes the polylogarithm)
\begin{eqnarray}
&&\sum_{k=1}^{n\ge 2} S(n,k)(k-1)! z^k = (-1)^n Li_{1-n}(1+1/z), \label{WS2-1}\\
&&\sum_{m=1}^n (-1)^m (m-1)! S(n,m) = 0\qquad {\rm for}\qquad n\ge 2. \label{WS2-2}
\end{eqnarray}

\subsection{Falling and rising factorials (Pochhammer symbols)}
\label{PS}

The falling factorial and rising factorials are defined as, correspondingly:
\begin{eqnarray}
(x)_n &=& x(x-1)(x-2)\cdots(x-n+1) = \frac{\Gamma(x+1)}{\Gamma(x-n+1)} = n!\binom{x}{n}, \label{PS-ff}\\
x^{(n)} &=& x(x+1)(x+2)\cdots(x + n -1) = \frac{\Gamma(x+n)}{\Gamma(x)} = n!\binom{x+n-1}{n}. \label{PS-rf}
\end{eqnarray}
In particular, $(x)_0 = 1$ and $x^{(0)} = 1$ because they are empty products (cf., Eqs. (\ref{def-ssnfk}) and (\ref{def-usstfk})). Note that the rising factorial can be expressed as a falling factorial that starts from the other end or as a falling factorial with opposite argument, correspondingly
\begin{equation}
x^{(n)} = ( x + n -1 )_n \qquad {or}\qquad x^{(n)} = (-1)^n(-x)_n. \label{ff-vs-rf}
\end{equation}

\section{Derivation of results for compound distribution $P_{CD} = P_{BD}\circ P_{PD}$}
\label{Derivation-1}

Using $F(z)$ and $G(z)$ for, respectively, BD and PD as given in Table \ref{Table_1} we have that (here and in the next Section $G^{(i)}(0) = G^{(i)}(z=0)$):
\begin{eqnarray}
&&F^{(r)}[G(0)] = r!\binom{K}{r}\left( p'\right)^r \left[ 1 - p' + p'G(0)\right]^{K-r} = r!X^K \binom{K}{r}Y^r; \label{F(k)-1}\\
&&L^{(r)}[G(0)] = (-1)^{r-1}(r-1)! K \left( p'\right)^r\left[ 1 - p' + p'G(0)\right]^{-r} =\nonumber\\
 &&\hspace{20mm} = (-1)^{r-1}(r-1)! K Y^r; \label{L(k)-1}\\
&&{\rm where} \qquad X = 1 - p' + p' e^{-\lambda}\qquad {\rm and}\qquad Y = \frac{p'}{X}e^{-\lambda},\label{Disc-BDPD-1}\\
&&\frac{G^{(m)}(z=0)}{m!} = \frac{\lambda^m e^{-\lambda}}{m!}, \label{G(m)-1}\\
&&\prod_{m=1}^{N-r+1} \left[ \frac{G^{(m)}(0)}{m!}\right]^{r_m} = \prod_{m=1}^{N-r+1} e^{-\lambda r_m} \prod_{m=1}^{N-r+1} \lambda^{m r_m} \prod_{m=1}^{N-r+1}\left( \frac{1}{m!} \right)^{r_m} = \nonumber\\
&&\hspace{35mm} = \lambda^N e^{-\lambda r} \prod_{m=1}^{N-r+1}\left[\frac{1}{m!}\right]^{r_m}. \label{G(m)-details}
\end{eqnarray}
Therefore the resulting multiplicity distribution $P(N)$ and the corresponding combinants $C^{\star}_N$ are :
\begin{eqnarray}
P(N) = \frac{1}{N!}\frac{d^N F(z)}{d z^N}\bigg|_{z=0}\! &=& \frac{1}{N!}\sum_{r=1}^N F^{(r)}\left[ G(0)\right]\cdot\nonumber\\
 &&\qquad\qquad \cdot B_{N,r}\left[ G^{(1)}(0),G^{(2)}(0),\dots,G^{N-r+1}(0)\right] =\nonumber\\
\! &=& \frac{\lambda^N}{N!}X^K \sum_{r=1}^{min(N,K)} r! \binom{K}{r} B_{N,r}(1,1,\dots,1)Y^r =\nonumber\\
&=& \frac{\lambda^N}{N!}X^K \sum_{r=1}^{min(N,K)} r! \binom{K}{r} S(N,r) Y^r; \label{PPD}\\
C^{\star}_N = \frac{1}{N!}\frac{d^N L(z)}{d z^N}\bigg|_{z=0}\! &=& K\frac{\lambda^N}{N!} \sum_{r=1}^N (-1)^{r-1}(r-1)! S(N,r) Y^r. \label{Cstar-BD-PD}
\end{eqnarray}
Note that in the BD summation over $r$ is restricted to $r\leq K$, consequently when calculating $P(N)$ the summation over $r$ is restricted to $min(N,K)$. The $B_{N,r}\left(\left\{x_i\right\}\right)$ are partial Bell polynomials defined by Eq. (\ref{Def-1}), their scaling property used here is given by Eq. (\ref{Def-3}) and their  relation with the Stirling numbers of the second kind, $S(N,r)$, is given in Eq. (\ref{Def-4}), while the Stirling numbers of the second kind are defined in Section \ref{Sn}.

Continuing, we note that because $S(N,0)=0$ (cf., Eq. (\ref{snfk-prop1})) the $P(0)$ calculated from Eq. (\ref{PPD}) is $P(0)=0$ instead of being equal (by definition) to $P(0) = F[G(z=0)] =  \left( 1 - p' + p'e^{-\lambda}\right)^K \label{P(0)}$. This is because the method of calculating derivatives of compound functions used here does not include (by definition) the $N=0$ derivative which must be added separately. In our case, this corresponds to starting the summation over $r$ from $r=0$ instead of $r=1$. Since from Section \ref{Sn} we know that $S(N,r) = 0$ for $r=0$ and for $r>N$, and $S(0,0)=1$, in this case we therefore have that  $P(0) = X^K\sum_{r=0}^K\binom{K}{r}S(0,r)Y^r e^{-r\lambda} = X^K = \left( 1 - p + pe^{-\lambda}\right)^K$, as expected. However, the result for the combinants given by Eq. (\ref{Cstar-BD-PD}) remains unchanged because $C^{\star}_N$ starts from $N=1$ and the problem with $N=0$ does not exist here.

The final result for $P(N)$ is therefore given by
\begin{equation}
P(N) = \frac{\lambda^N}{N!} X^K \sum_{r=0}^{min(N,K)} r! \binom{K}{r} S(N,r) Y^r. \label{PPD-restr}
\end{equation}
Eqs. (\ref{PPD-restr}) and (\ref{Cstar-BD-PD}) represent our final result for the  multiplicity distribution and combinants for a compound distribution $P_{CD} = P_{BD}\circ P_{PD}$. If necessary, it can be expanded further by using the representation of  $S(N,r)$ given by Eq. (\ref{S1-a}) in Section \ref{Sn}, namely that $(-1)^r r!S(N,r) = \sum_{l=0}^r \binom{r}{l}(-1)^l l^N$.

This result can also be shown in another way, in which the combined distribution structure is better visible. For this, let us define first the parameter $\tilde{p}$  such that
\begin{equation}
\tilde{p} = \frac{p'}{1+p' e^{-\lambda}} \label{tildep}
\end{equation}
and  write Eq. (\ref{PPD-restr}) as follows:
\begin{eqnarray}
P(N) &=& \frac{\lambda^N}{N!} X^K \sum_{r=0}^{min(N,K)} r! \binom{K}{r} S(N,r) Y^r = \nonumber\\
&=& \frac{\lambda^N}{N!}\sum_{r=0}^{min(N,K)}\binom{K}{r}\left[ r! e^{-r\lambda} S(N,r)\right]\cdot \left( 1 - p' + p'e^{-\lambda}\right)^{K-r} \left( p'\right)^r =\nonumber\\
&=& \frac{\lambda^N}{N!}\left( 1 + p'e^{-\lambda}\right)^K\sum_{r=0}^{min(N,K)} P_{BD}(r)\cdot \left[ r! e^{-r\lambda} S(N,r)\right], \label{P_r}\\
P(0) &=& \left( 1 + p' e^{-\lambda}\right)^K P_{BD}(0)S(0,0) = \left( 1 - \tilde{p}\right)^K\left( 1 - p' + p' e^{-\lambda} \right)^K = \nonumber\\
&=& \left( 1 - p' + p'e^{-\lambda}\right)^K, \label{P_r0}
 \end{eqnarray}
where $P_{BD}(r) =  \binom{K}{r}\left( \tilde{p}\right)^r \left( 1 - \tilde{p}\right)^{K-r}$, i.e., it has the form characteristic for the BD multiplicity distribution listed in Table \ref{Table_1} with $p \to \tilde{p}$. Similarly, Eq. (\ref{Cstar-BD-PD}) can be rewritten as
\begin{eqnarray}
C^{\star}_N &=& K\frac{\lambda^N}{N!} \sum_{r=1}^N (-1)^{r-1} (r-1)! S(N,r)Y^r =\nonumber\\
 &=& \frac{\lambda^N}{N!} \sum_{r=1}^N (-1)^{r-1} \frac{K}{r}r! e^{-r\lambda} S(N,r)\left( \frac{p'}{1-p'+p'e^{-\lambda}} \right)^r = \nonumber\\
 &=&  \frac{\lambda^N}{N!} \sum_{r=1}^N C^{\star}_r(BD)\cdot r! e^{-r\lambda} S(N,r), \label{C=CBDfactor}
\end{eqnarray}
where $C^{\star}_r = (-1)^{r-1}\frac{K}{r}\left(\frac{\tilde{p}}{1-\tilde{p}}\right)^r$, i.e., it has  the form characteristic for the combinants for a BD multiplicity distribution listed in Table \ref{Table_1} in which parameter $p$ is replaced by $\tilde{p}$. It is straightforward to check that for $K=1$ and $p'=1$ (we only have a single object producing particles distributed according to a PD) the above equations (\ref{P_r}) and (\ref{C=CBDfactor})  describe, respectively, $P(N)$ and $C^{\star}_N$ for the Poisson distribution.

\section{Derivation of results for compound distribution $P_{CD} = P_{BD}\circ P_{NBD}$}
\label{Derivation-2}

Using $F(z)$ and $G(z)$ for, respectively, the BD and PD as given in Table \ref{Table_1} we have that:
\begin{eqnarray}
&&F^{(r)}[G(0)] = r!\binom{K}{r}\left( p'\right)^r \left[ p'z + 1 - p'G(0)\right]^{K-r} = r!\tilde{X}^K \binom{K}{r} \tilde{Y}^r; \label{F(k)-2}\\
&&L^{(r)}[G(0)] = (-1)^{r-1}(r-1)! K \left( p'\right)^r\left[ 1 - p' + p'G(0)\right]^{-r} = \nonumber\\
&& \hspace{20mm} = (-1)^{r-1}(r-1)! K \tilde{Y}^r; \label{L_(k)-2}\\
&& {\rm where}\qquad
\tilde{X} = 1 - p' + p' (1 - p)^k\qquad{\rm and}\qquad \tilde{Y} = \frac{p'(1-p)^k}{\tilde{X}}, \label{Def-tildeXY}\\
&&\frac{G^{(m)}(0)}{m!} = p^m \binom{k+m-1}{m} (1-p)^k; \label{G(m)-2}
\end{eqnarray}
\begin{eqnarray}
&&\prod_{m=1}^{N-r+1} \left[ \frac{G^{(m)}(0)}{m!}\right]^{r_m} = \prod_{m=1}^{N-r+1}\left[p^m\binom{k+m-1}{m}(1-p)^k \right]^{r_m} = \nonumber\\
&&\hspace{35mm} = p^N(1-p)^{kr}\prod_{m=1}^{N-r+1}\left[\frac{k^{(m)}}{m!}\right]^{r_m},  \label{prodG}
\end{eqnarray}
where $k^{(m)} = \prod_{i=0}^{m-1}(k+i) = k(k+1)\cdots(k+m-1)$ is a rising factorial defined in Eq. (\ref{PS-rf}). Defining
\begin{eqnarray}
&& B_{N,r}\left[k^{(1)},k^{(2)},\dots,k^{(N-r+1)}\right] = B_{N,r}\left[ \left\{ k^{(N-r+1)}\right\}\right], \label{defBNr}\\
&& B_{N,r}\left[ G^{(1)}(0),G^{(2)}(0),\dots,G^{(N-r+1)}(0)\right] = B_{N,r}\left[ \left\{ G^{(N-r+1)}(0)\right\}\right], \label{defBNr1}
\end{eqnarray}
we have that
\begin{eqnarray}
&&P(N)=\frac{1}{N!}\frac{d^N F(z)}{d z^N}\bigg|_{z=0} = \frac{1}{N!}\sum_{r=1}^N F^{(r)}\left[ G(0)\right]B_{N,r}\left[ \left\{ G^{(N-r+1)}(0)\right\}\right] =\nonumber\\
&&\qquad = \frac{p^N}{N!}\tilde{X}^K  \sum_{r=1}^{min(N,K)} r!\binom{K}{r} B_{N,r}\left[ \left\{ k^{(N-r+1)}\right\}\right]\cdot \tilde{Y}^r,  \label{BD-NBD}\\
&&C^{\star}_N = \frac{1}{N!}\frac{d^N L(z)}{d z^N}\bigg|_{z=0} = K \frac{p^N}{N!} \sum_{r=1}^N (-1)^{r-1} (r-1)! B_{N,r}\left[ \left\{ k^{(N-r+1)}\right\}\right]\cdot \tilde{Y}^r. \label{Cstar-BD-NBD-fin}
\end{eqnarray}
For the same reasons as in Section \ref{CD=BDxPD} when we calculate $P(N$) here  the summation over $r$ is limited to $r \leq min(N,K)$ and because originally  the $P(0)$ term is missing in  Eqs. (\ref{BD-NBD}) it must be added. This is equivalent to a change of summation in the inner sum from $r=1$ to $r=0$. However, the $C^{\star}_N$ are not affected and remain unchanged. The final result for $P(N)$ is therefore given by
\begin{equation}
P(N) = \frac{p^N}{N!}\tilde{X}^K \cdot \sum_{r=0}^{min(N,K)} r!\binom{K}{r} B_{N,r}\left[ \left\{ k^{(N-r+1)}\right\}\right]\cdot \tilde{Y}^r.\label{BD-NBD-final-1}
\end{equation}

Eqs. (\ref{BD-NBD-final-1}) and (\ref{Cstar-BD-NBD-fin}) represent our final result for the multiplicity distribution and combinants for the compound distribution $P_{CD} = P_{BD}\circ P_{NBD}$. In fact, we can go further in analytic calculations by replacing the Bell plynomials there with their analytical expressions,
 $$B_{n,r}\left[x^{(1)},x^{(2)},\dots,x^{(n-r+1)} \right] = \frac{(-1)^r}{r!}\sum_{l=0}^r (-1)^l \binom{r}{l}(xl)^{(n)},$$
given by Eq. (\ref{Def-8}) from Section \ref{BP} with the rising factorial $(x\cdot l)^{(n)}$ given by Eq. (\ref{PS-rf}). Note that according to Eqs. (\ref{PS-ff}) and (\ref{PS-rf}) $(x)_n =0$ for $x=0$ and, similarly, $x^{(n)}=0$ for $x=0$. Therefore, when inserted in Eqs. (\ref{BD-NBD-final-1}) and (\ref{Cstar-BD-NBD-fin}) the inner summations over $l=0,\dots,r$ start in reality from $l=1$. In particular this  means that the first term in the inner sum in Eq. (\ref{Cstar-BD-NBD-fin}) is always positive.

Also in this case our result can be presented in a way clearly exposing the essence of the compound distribution $P_{CD} = P_{BD}\circ P_{NBD}$. Let us introduce parameter $\hat{p}$  such that
\begin{equation}
\hat{p} = \frac{p'}{1+p' (1-p)^k}. \label{tildepNBD}
\end{equation}
Eq. (\ref{BD-NBD-final-1})  can now be written in the following way:
\begin{eqnarray}
&&P(N) = \frac{p^N}{N!}\left[ 1 - p' + p' (1 - p)^k\right]^K  \cdot \nonumber\\
&&\hspace{25mm} \cdot \sum_{r=0}^{min(N,K)} r!\binom{K}{r} B_{N,r}\left[ \left\{ k^{(N-r+1)}\right\}\right]\left[ \frac{ p' (1 - p)^k }{ 1 - p' + p' (1 - p)^k}\right]^r = \nonumber\\
&&\qquad = \frac{p^N}{N!}\sum_{r=0}^{min(N,K)} r!\binom{K}{r} (1-p)^{kr}\cdot\nonumber\\
&&\hspace{25mm} \cdot B_{N,r}\left[ \left\{ k^{(N-r+1)}\right\}\right]\left[1 - p' + p' (1 - p)^k\right]^{K-r}\left( p'\right)^r = \nonumber\\
&&\qquad = \frac{p^N}{N!}\left[ 1 + p'(1-p)^k\right]^K \cdot\nonumber\\
&&\hspace{25mm} \cdot \sum_{r=0}^{min(N,K)} P_{BD}(r) \cdot\left\{ r! (1-p)^{rk}B_{N,r}\left[ \left\{ k^{(N-r+1)}\right\}\right] \right\} \label{Pr-BD*NBD};\\
&&P(0) = \left[ 1 + p'(1-p)^k\right]^K P_{BD}(r) = \nonumber\\
&& \qquad = \left[ 1 + p'(1-p)^k\right]^K \left( 1 - \hat{p}\right)^K = \left[ 1 -p' + p'(1-p)^k\right]^K, \label{Pr-BD*NBD0}
\end{eqnarray}
where $P_{BD}(r) =  \binom{K}{r}\left( \hat{p}\right)^r \left( 1 - \hat{p}\right)^{K-r}$ in the $r$-th component in the total of $P(N)$ has the form characteristic for the BD multiplicity distribution, as given in Table \ref{Table_1} with $p\to \hat{p}$. Accordingly, Eq. (\ref{Cstar-BD-NBD-fin}) can be written as:
\begin{eqnarray}
C^{\star}_N &=& K \frac{p^N}{N!} \cdot \sum_{r=1}^N (-1)^{r-1} (r-1)! B_{N,r}\left[ \left\{ k^{(N-r+1)}\right\}\right]\cdot \left[ \frac{ p' (1 - p)^k}{ 1 - p' + p' (1 - p)^k}\right]^r =\nonumber\\
 &=& \frac{p^N}{N!} \sum_{r=1}^N \left[ (-1)^{r-1} \frac{K}{r}\left(\frac{\hat{p}}{1-\hat{p}}\right)^r\right]\cdot r! (1-p)^{rk} B_{N,r}\left[ \left\{ k^{(N-r+1)}\right\}\right] = \nonumber\\
&=& \frac{p^N}{N!} \sum_{r=1}^N C^{\star}_r(BD)\cdot r! (1-p)^{rk} B_{N,r}\left[ \left\{ k^{(N-r+1)}\right\}\right], \label{C=CBDNBDfactor}
\end{eqnarray}
where the $C^{\star}_r(BD) = (-1)^{r-1} \frac{K}{r}\left(\frac{\hat{p}}{1-\hat{p}}\right)^r$ in the $r$-th component in the total of $C^{\star}_N$ has the form characteristic for the BD as given in Table \ref{Table_1} in which parameter $p'$ is replaced by $\hat{p}$.

Note that in the limit $k \to \infty$ with $\langle N\rangle = const$, when the NBD turns into the PD, one has that $ (1 - p)^k \to e^{-\langle N\rangle}$ and $(1-p)^{rk} \to e^{-r\langle N\rangle}$, this means that $\hat{p} \Longrightarrow  \tilde{p}$   and
\begin{eqnarray}
\frac{p^N}{N!}\cdot B_{N,r}\left[ \left\{ k^{(N-r+1)}\right\}\right] &\Longrightarrow & \frac{\lambda^N}{N!}\cdot \left\{\left( \frac{1}{k}\right)^N B_{N,r}\left[ \left\{ k^{(N-r+1)}\right\}\right]\right\} \to \nonumber\\
&\to& \frac{\lambda^N}{N!} S(N,r), \label{X2}
\end{eqnarray}
(where we have used Eqs. (\ref{Def-3}), (\ref{PS-rf}) and (\ref{Def-4})). So in this limit Eqs. (\ref{Pr-BD*NBD}) and (\ref{C=CBDNBDfactor}) become Eqs. (\ref{P_r}) and (\ref{C=CBDfactor}), and $P_{BD}\circ P_{NBD}$ becomes $P_{BD}\circ P_{PD}$. Similarly as in the Section \ref{CD=BDxPD} one can check that for $K=1$ and $p'=1$ one again reproduces $P(N)$ and $C^{\star}_N$ for the NBD.

\end{document}